%% file: city_paper.tex
\documentclass[final,1p,times]{elsarticle}

\usepackage{amssymb}
\usepackage{amsmath}
\usepackage{subfig}
\usepackage{graphicx}
\usepackage{pgfplots}
\usepackage{pgfplotstable}
\usepackage{bm}
\usepackage{filecontents}
\usepackage{url}
\usepackage{tikz}
\usepackage{overpic}
\usepackage{multirow}
\usepackage{lscape}
\usepackage{booktabs}

\newcommand{\windrose}[3][black]{%
    \begin{tikzpicture}[scale=#3, transform shape] %
        \draw[thick, color=#1] (0,-1) -- (0,1);
        \draw[thick, color=#1] (-1,0) -- (1,0);

        \node[color=#1, anchor=south] at (0,1.2) {\normalsize \textbf{N}};
        \node[color=#1, anchor=north] at (0,-1.2) {\normalsize \textbf{S}};
        \node[color=#1, anchor=east] at (-1.2,0) {\normalsize \textbf{W}};
        \node[color=#1, anchor=west] at (1.2,0) {\normalsize \textbf{E}};

        \draw[thick,->,color=#1] (0,0) -- ({cos(#2)}, {sin(#2)});
    \end{tikzpicture}
}

\usepackage{lineno}
\journal{Building and Environment}

\begin{document}

\begin{frontmatter}

\title{A Digital Urban Twin Enabling Interactive Pollution Predictions and Enhanced Planning} %

\author[1,2]{Dennis Teutscher}
\author[1,2]{Fedor Bukreev}
\author[1,4]{Adrian Kummerländer}
\author[1,4]{Stephan Simonis}
\author[2]{Peter Bächler}
\author[3]{Ashkan Rezaee}
\author[3]{Mariusz Hermansdorfer}
\author[1,2,4]{Mathias J.\ Krause}
\affiliation[1]{organization={Lattice Boltzmann Research Group, Karlsruhe Institute of Technology},%
    addressline={Englerstr.~2}, 
    city={Karlsruhe},
    postcode={76131}, 
    state={Baden-Württemberg},
    country={Germany}}
\affiliation[2]{organization={Institute for Mechanical Process Engineering and Mechanics, Karlsruhe Institute of Technology},%
    addressline={Straße am Forum~8}, 
    city={Karlsruhe},
    postcode={76131}, 
    state={Baden-Württemberg},
    country={Germany}}
\affiliation[4]{organization={Institute for Applied and Numerical Mathematics, Karlsruhe Institute of Technology},%
    addressline={Englerstr.~2}, 
    city={Karlsruhe},
    postcode={76131}, 
    state={Baden-Württemberg},
    country={Germany}}
\affiliation[3]{organization={Henning Larsen},%
    addressline={Vesterbrogade 76}, 
    city={Copenhagen},
    postcode={1620}, 
    state={Hovedstaden},
    country={Denmark}}

\begin{abstract}
Digital twin (DT) technology is increasingly used in urban planning, leveraging real-time data integration for environmental monitoring. This paper presents an urban-focused DT that combines computational fluid dynamics simulations with live meteorological data to analyze pollution dispersion. Addressing the health impacts of pollutants like particulate matter and nitrogen dioxide, the DT provides real-time updates on air quality, wind speed, and direction.
Using OpenStreetMap’s XML-based data, the model distinguishes between porous elements like trees and solid structures, enhancing urban flow analysis. The framework employs the lattice Boltzmann method (LBM) within the open-source software OpenLB to simulate pollution transport. Nitrogen dioxide and particulate matter concentrations are estimated based on traffic and building emissions, enabling hot-spot identification.
The DT was used from November 7 to 23, 2024, with hourly updates, capturing pollution trends influenced by wind patterns. Results show that alternating east-west winds during this period create a dynamic pollution distribution, identifying critical residential exposure areas.
This work contributes a novel DT framework that integrates real-time meteorological data, OpenStreetMap-based geometry, and high-fidelity LBM simulations for urban wind and pollution modeling. Unlike existing DTs, which focus on structural monitoring or large-scale environmental factors, this approach enables fine-grained, dynamic analyses of urban airflow and pollution dispersion. By allowing interactive modifications to urban geometry and continuous data updates, the DT serves as a powerful tool for adaptive urban planning, supporting evidence-based policy making to improve air quality and public health.
\end{abstract}

\begin{graphicalabstract}
\includegraphics[width=\textwidth]{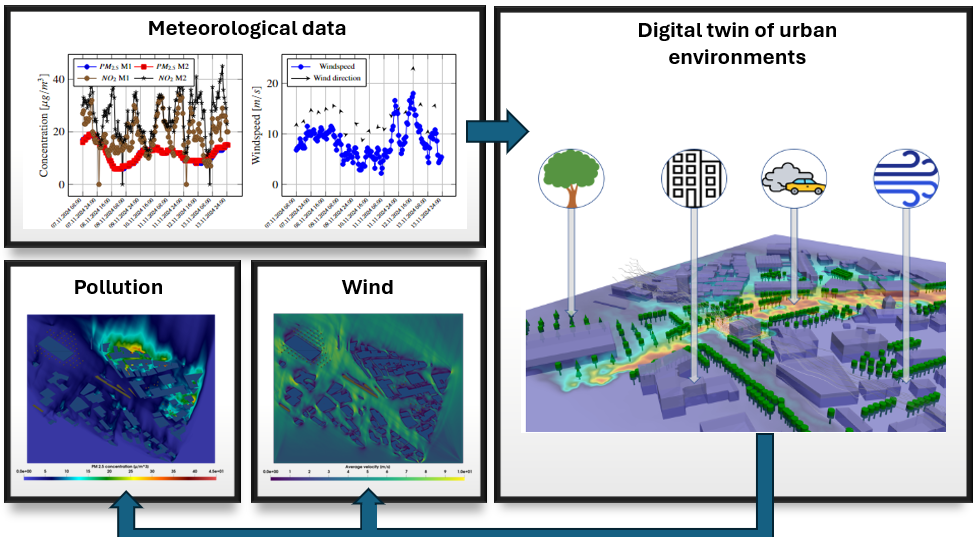}
\end{graphicalabstract}

\begin{highlights}
    \item Digital twin model that uses the homogenized lattice Boltzmann method (HLBM) for efficient urban airflow and pollution simulations.
    \item Integrated real-time meteorological data to dynamically update wind and air pollution distributions, enhancing the realism of simulations
    \item Identifying pollution hot-spots by analyzing long-term data trends.
    \item Interactive urban geometry adjustments using OpenStreetMap, allowing flexible adaptation to evolving city structures and planning scenarios.
\end{highlights}


\end{frontmatter}


\section{Introduction}
With the emergence of Industry 4.0 and the impending Industry 5.0, the term digital twin (DT) has become a cornerstone in discussions about future technological advancements. The concept of DT was first introduced by Grieves~\textit{et al.} in 2002~\cite{GRIEVES2017}. DTs are already being utilized in various fields, such as manufacturing, where they represent the life cycle of products~\cite{TAO18,ZHANG21}; aerospace, where they monitor plane condition to improve reliability and safety~\cite{GLAESSGEN12}; and business planning, where self-evolving DTs optimize business models~\cite{KLOSTERMEIER19}. In addition, DTs are increasingly being used in infrastructure and building planning, as well as in hazard mitigation. They can monitor the stress distribution in structures~\cite{FACUNDO23} and oversee the life cycle of bridges~\cite{HONGHONG23,GAO23,YE19}.

Although DTs are essential for monitoring structural stability and detecting damage, they must also account for external factors such as air pollution and high wind speeds in narrow street canyons. Recent advances have been made in the creation of DTs for cities. Schrotter~\textit{et al.}~\cite{SCHROTTER20} presented a DT for urban planning that monitors environmental factors such as noise and temperature.

Beyond DTs, computational fluid dynamics (CFD) simulations have been extensively used to model urban environments. Pasquier~\textit{et al.}~\cite{PASQUIER23} simulated the traffic emission distribution in a city using the lattice Boltzmann method (LBM) with the open-source software \textit{OpenLB}~\cite{OpenLB,OLBRELEASE,USERGUIDE} and validated the model. Van Hooff~\textit{et al.}~\cite{VANHOOFF10} simulated air flow around urban structures. Jeanjean~\textit{et al.}~\cite{JEANJEAN15} studied the effects of trees on pollution dispersion in cities, using a wind tunnel for validation. Taleb~\textit{et al.}~\cite{TALEB21} investigated how trees function as windshields to lower the concentration of dust particles in desert cities.
The relevance of this research lies in the adverse health effects caused by air pollution, e.g., through exposure to particulate matter~\cite{KIM15} and gaseous substances such as nitrogen dioxide~\cite{BONINGARI16}, which impact respiratory and cardiovascular health. Monitoring the pollution distribution (e.g., by government-official monitoring sites) and integrating this data into urban planning is crucial for public health. Within the revision of the EU Ambient Air Quality Directive, monitoring of non-regulated "new pollutants" such as ultra fine particles or black carbon is becoming mandatory. Depending on the size and the number of inhabitants of the member state, there is an obligation to install a number of "supersites" that monitor these pollutants. However, ultra fine particles and other pollutant concentrations are known to fluctuate and local exposure depends strongly on the transmission conditions as well as the spatial and temporal behavior of pollutant sources~\cite{DALL'OSTO13}. The challenges arise from the complexity of identifying the sources of pollution and their impact on local ambient concentration levels~\cite{GARCIA-MARLÈS24}. For example, Dröge~\textit{et al.}~\cite{DRÖGE24} showed the significance of pollutant transmission for ultra fine particle concentrations from air traffic and the high temporal variations. Other prominent urban sources of (ultra)fine particles and air pollution are traffic~\cite{MOHAN24} and public transport~\cite{SAMAD22}. Furthermore, heating from wood stoves and other combustion processes, such as barbecues, can cause a significant increase in particle concentration and gaseous pollutants, especially in densely populated (residential) areas~\cite{BAECHLER24}. Spatially resolved measurements to investigate local particle concentrations are associated with a high amount of measurement effort~\cite{BAECHLER21} or limited accuracy considering the use of distributed low-cost sensors~\cite{KAUR23}. Therefore, simulations are an excellent tool to predict pollutant transmission and determine local concentration levels in the context of limited numbers of monitoring sites.
Although previous studies have leveraged DTs for urban applications, they have focused primarily on structural integrity or large-scale environmental factors such as noise. However, a DT specifically designed to model wind distribution and air pollution in urban areas, while integrating real-time meteorological data and enabling interactive geometry adjustments using OpenStreetMap (OSM)~\cite{OPENSTREETMAP}, remains largely unexplored.
To address these gaps, this paper presents a novel DT that:
\begin{itemize}
    \item Utilizes the homogenized lattice Boltzmann method (HLBM) for accurate and computationally efficient fluid simulations in urban environments,
    \item Integrates real-time meteorological data to dynamically update wind and air pollution distributions, enhancing the realism of simulations,
    \item Enables interactive urban geometry adjustments using OSM, allowing flexible adaptation to evolving city structures and planning scenarios.
\end{itemize}
By combining these elements, our DT framework extends beyond traditional static urban models, providing a dynamic and interactive tool for air quality monitoring and urban planning.
The remainder of this paper is structured as follows: Section~\ref{sec:meth} outlines the methodology, including the DT concept, integration of OSM data, mathematical modeling, and discretization using HLBM. Section~\ref{sec:validation} presents the validation of the numerical approach using experimental data. Finally, Section~\ref{sec:res} discusses the DT results, followed by the conclusions in Section~\ref{sec:conc}.

\section{Methodology}\label{sec:meth}

\subsection{Concept of digital twin for urban areas}

\begin{figure}[h]
    \centering
    \includegraphics[width=0.8\linewidth]{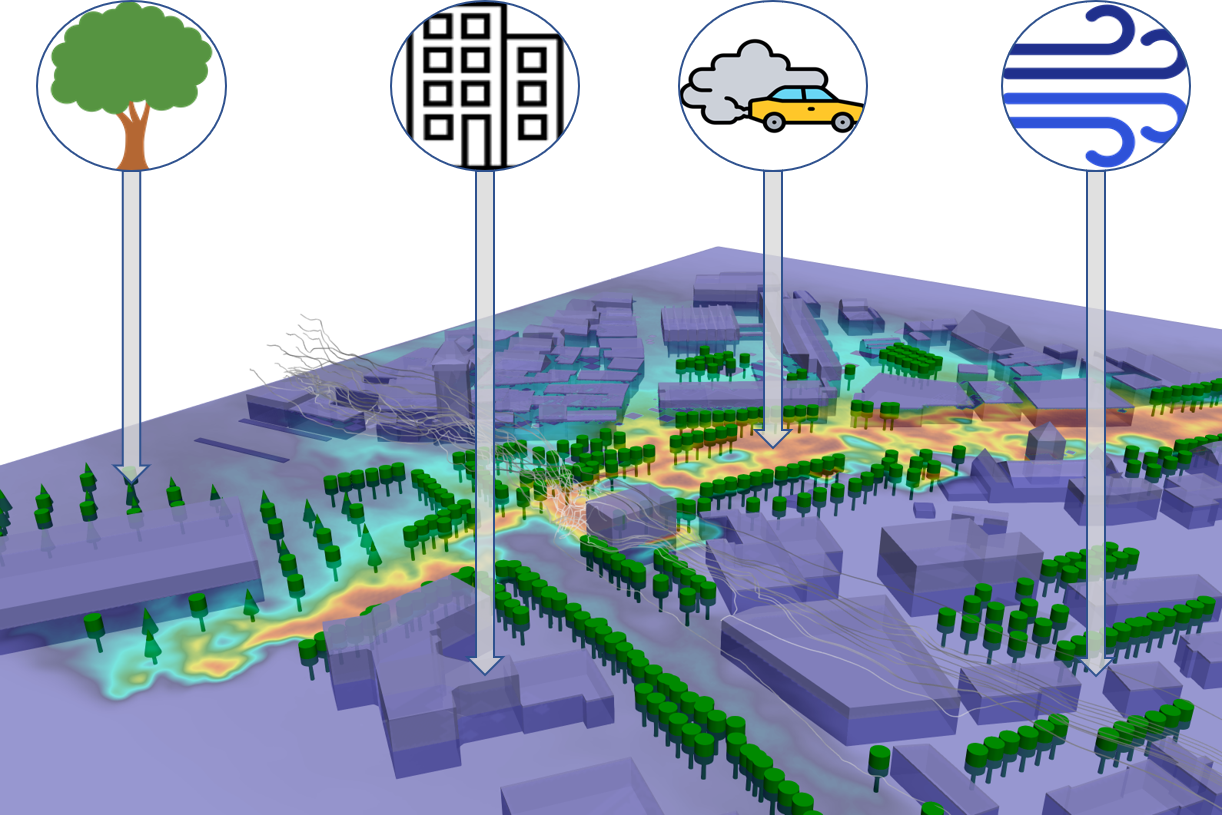}
    \caption{Conceptual visualization of the Digital Twin for an urban environment, illustrating key data elements such as buildings and porous zones (Section~\ref{sec:geo_creation}), pollution sources including vehicles and buildings (Section~\ref{sec:sources}), as well as wind speed and direction, which dynamically update a numerical model (Section~\ref{sec:modelling}).}  
    \label{fig:concept}
\end{figure}

The proposed DT concept, illustrated in Figure~\ref{fig:concept}, is centered on a simulation framework with two primary interfaces. The first interface enables automatic updates to the simulation using real-time data from measuring stations, while the second is a manual interface that allows users to adjust the geometry in the simulation, integrating and removing data from OSM into the simulation.
Creating a DT for urban environments requires the consideration of multiple dynamic factors, especially when the objective is to assist urban planners and policymakers in making data-driven decisions. One critical application of this DT is to assess the distribution of pollutants, such as particulate matter and nitrogen dioxide, throughout the city. Identifying sources of pollution poses significant challenges, especially with particulate matter, which originates from various sources such as traffic, industrial emissions, and domestic heating. This complexity is further compounded by the limitations of current measurement infrastructure, as most urban measuring stations provide only localized point-based data.
This concept is based on the idea to give city planners and policy makers the ability to insert, for example, a new pollution source into the area and manipulate the distribution by inserting trees and buildings. 
To use it as a planning tool, it is significant that the performance is good enough to give reasonable, quick results. For that reason, we use the open-source software \textit{OpenLB} which allows the use of GPUs and is highly scaleable by using more GPUs.

\subsection{Interactive data input}
To ensure that the DT accurately reflects the state of its physical counterpart, the urban area, there must be a constant exchange of data. For this purpose, we utilize publicly available measuring stations that provide real-time information on wind speed, wind direction, and the concentrations of particulate matter and nitrogen dioxide. This data is then used to update the DT at regular intervals, showcasing the current state of the monitored area. By averaging the results over a longer period, pollution hot spots are identified.

\subsection{OpenStreeMap}
In addition to data from measuring stations, information about the types of buildings and objects in the area is also crucial. For this, we use the open-source map OSM, which stores data in an XML-based format. Each object is marked with a specification of what kind of object it is, allowing us to differentiate between porous objects such as trees and solid structures. Since the data is XML-based, it is easy to add or remove information, making it suitable for planning urban areas and exploring different scenarios.

\subsubsection{Parsing OSM data}
The first step involves reading and interpreting OSM XML files using the TinyXML2~\cite{TINYXML2} library. The nodes and ways, which are the fundamental elements of OSM, are parsed for relevant geographical data. 

\paragraph{Nodes parsing}
Nodes represent specific points with latitude and longitude. These are extracted as key-value pairs using attributes like lat and lon.
The nodes are stored in an unordered\_map to allow a quick lookup by their unique identifiers (IDs).
If a node represents a tree, additional tags such as leaf\_type and height are parsed. Tags are analyzed using a function that checks the presence of natural tags like tree, wood, or scrub.

\paragraph{Ways parsing}
Ways are sequences of nodes that define streets, buildings, or other linear/area features. Attributes for specific way types are filtered.
Buildings are identified using a predefined set of building-related tags (e.g.\ residential, commercial) and are represented as a list of their footprint nodes.
Roads are parsed by their name and width when available.
Trees and natural features are parsed similarly, focusing on attributes like height or species. \\

\begin{figure}[ht]
    \centering
    \begin{tikzpicture}[node distance=2cm, auto]
        \node (osm) [rectangle, draw, fill=blue!20] {OSM XML file};
        \node (node) [rectangle, draw, fill=green!20, below of=osm, xshift=-3cm] {\textless node\textgreater{} elements};
        \node (way) [rectangle, draw, fill=green!20, below of=osm, xshift=3cm] {\textless way\textgreater{} elements};
        
        \node (parseNodes) [rectangle, draw, fill=yellow!20, below of=node] {Parse nodes};
        \node (storeNodes) [rectangle, draw, fill=yellow!20, below of=parseNodes] {Store in unordered\_map};
        
        \node (parseWays) [rectangle, draw, fill=yellow!20, below of=way] {Parse ways};
        \node (categorizeWays) [rectangle, draw, fill=yellow!20, below of=parseWays] {Categorize as buildings/streets/ trees};
        
        \node (buildings) [rectangle, draw, fill=red!20, below of=storeNodes, xshift=-2cm] {Buildings vector};
        \node (streets) [rectangle, draw, fill=red!20, below of=storeNodes, xshift=2cm] {Streets vector};
        
        \draw[->] (osm) -- (node);
        \draw[->] (osm) -- (way);
        \draw[->] (node) -- (parseNodes);
        \draw[->] (parseNodes) -- (storeNodes);
        \draw[->] (way) -- (parseWays);
        \draw[->] (parseWays) -- (categorizeWays);
        \draw[->] (storeNodes) -- (buildings);
        \draw[->] (storeNodes) -- (streets);
    \end{tikzpicture}
    \caption{Parsing workflow of OSM data, illustrating the extraction and categorization of nodes and ways into buildings and streets.}
    \label{fig:parsing_process}
\end{figure}
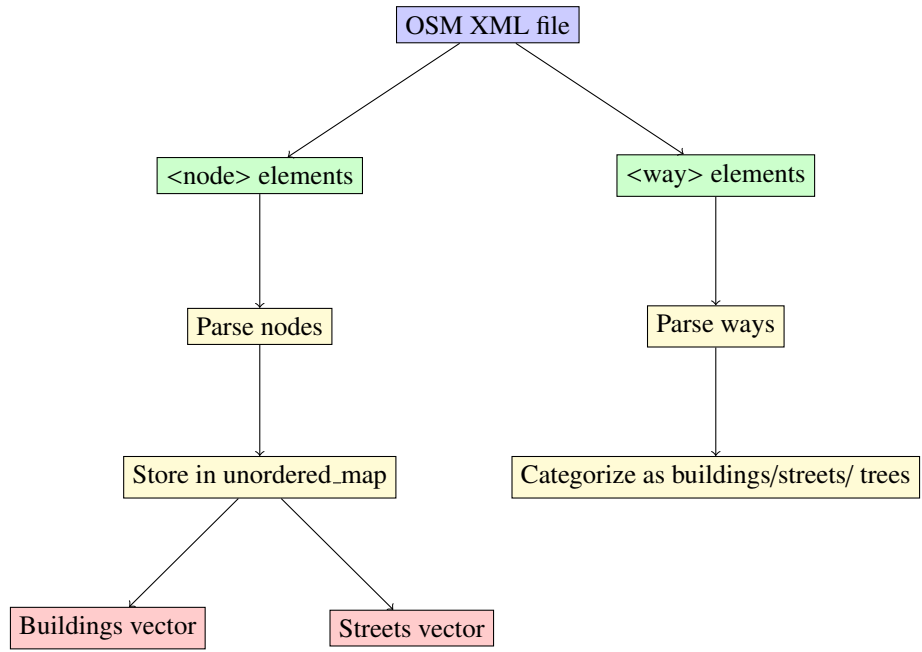

Figure~\ref{fig:parsing_process} illustrates the parsing process for nodes and ways, highlighting the extraction of essential attributes and relationships between OSM elements.

\subsubsection{Data transformation and filtering}\label{sec:osm_data}
Parsed data undergoes transformation to prepare it for 3D geometry creation or visualization.

\paragraph{Building data transformation}
Building footprints are transformed into local UTM (Universal Transverse Mercator) coordinates:
The UTM zone is dynamically determined on the basis of the longitude of the first building node.
Using the PROJ library, all geographical coordinates are converted into projected coordinates relative to the calculated UTM zone.
To ensure compatibility with local systems, a translation is applied to shift coordinates into a local reference frame.

\paragraph{Tree and natural feature processing}
Trees are processed similarly, with special attention to their height and species. If height is not explicitly defined, default values are assigned to avoid anomalies. Tree rows and areas (e.g., forest, scrubs) are treated as collections of nodes.

\subsubsection{Geometry creation}\label{sec:geo_creation}
The processed data is used to create 3D geometries that represent the extracted OSM elements.

\begin{figure}[ht]
    \centering
    \begin{tikzpicture}[scale=1.0]
        \fill[gray!30] (0,0) -- (4,0) -- (3,2) -- (1,2) -- cycle; %
        \draw[thick] (0,0) -- (4,0) -- (3,2) -- (1,2) -- cycle; %
        \node at (2, -0.5) {2D building footprint};

        \draw[dashed, ->] (0,0) -- (0,4) node[midway, left] {Height};
        \draw[dashed, ->] (4,0) -- (4,4);
        \draw[dashed, ->] (1,2) -- (1,6);
        \draw[dashed, ->] (3,2) -- (3,6);

        \begin{scope}[shift={(6,0)}]
            \fill[gray!30] (0,0) -- (4,0) -- (3,4) -- (1,4) -- cycle; %
            \fill[gray!50, opacity=0.7] (4,0) -- (4,4) -- (3,6) -- (3,2) -- cycle;
            \fill[gray!30, opacity=0.5] (0,4) -- (1,6) -- (3,6) -- (4,4) -- cycle; 
            \fill[gray!40, opacity=0.7] (0,0) -- (1,2) -- (1,6) -- (0,4) -- cycle; 
            
            \draw[thick] (0,0) -- (4,0) -- (3,4) -- (1,4) -- cycle; %
            \draw[thick] (4,0) -- (4,4); %
            \draw[thick] (3,4) -- (3,6); %
            \draw[thick] (1,4) -- (1,6); %
            \draw[thick] (0,4) -- (4,4); %
            \draw[thick] (0,0) -- (0,4); %
            \draw[thick] (1,6) -- (3,6);
            \draw[thick] (0,4) -- (1,6);
            \draw[thick] (4,4) -- (3,6);
            \node at (2, 7) {3D extruded building};
        \end{scope}
    \end{tikzpicture}
    \caption{Extrusion of a 2D building footprint into a 3D volume, starting from the ground up, based on building height.}
    \label{fig:building_extrusion}
\end{figure}
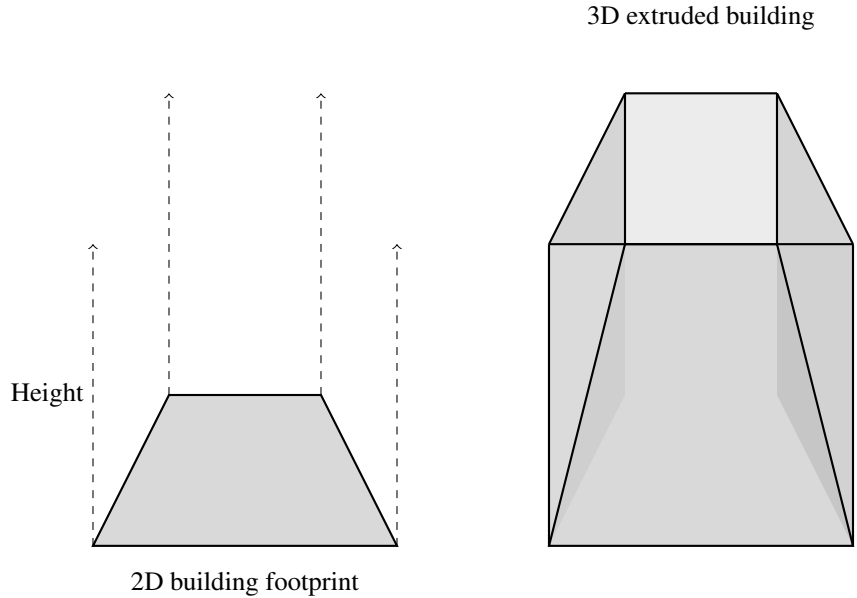

\paragraph{Buildings}
Building geometries are constructed as 3D prisms, with footprints that define the base and the height determined by the parsed or default values. Indicator polygons are created using the IndicatorPolygon3D class to encapsulate the footprint points. Figure~\ref{fig:building_extrusion} shows the footprint of the sample building extruded into a 3D volume.

\paragraph{Streets}
Street geometries are expanded to 3D surfaces based on their width. A defined width or default value (e.g.\ 5 meters) is applied to calculate a polygonal representation of the street area, ensuring complete coverage beyond simple outlines.

\begin{figure}[ht]
    \centering
    \begin{tikzpicture}[scale=1.0]
        \begin{scope}[shift={(-3,0)}]
            \draw[thick, fill=brown!70] (0,0) ellipse (0.2 and 0.1); %
            \draw[thick, fill=brown!70] (0,2) ellipse (0.2 and 0.1); %
            \draw[thick, fill=brown!70] (-0.2,0) -- (-0.2,2); %
            \draw[thick, fill=brown!70] (0.2,0) -- (0.2,2); %

            \draw[thick, fill=green!50] (-1,2) rectangle (1,4); %
            \draw[thick, fill=green!50] (0,4) ellipse (1 and 0.5); %
            \draw[thick, fill=green!50] (0,2) ellipse (1 and 0.5); %

            \draw[thick] (-1,2) -- (-1,4); %
            \draw[thick] (1,2) -- (1,4); %

            \node at (0, 5.5) {3D Tree Representation};
        \end{scope}

        \begin{scope}[shift={(4,0)}]
            \fill[green!30] (0,0) -- (3,0) -- (2,2) -- (1,2) -- cycle; %
            \draw[thick] (0,0) -- (3,0) -- (2,2) -- (1,2) -- cycle; %

            \fill[green!30, opacity=0.5] (0,0) -- (1,2) -- (1,4) -- (0,2) -- cycle; %
            \fill[green!30, opacity=0.5] (3,0) -- (2,2) -- (2,4) -- (3,2) -- cycle; %
            \fill[green!30] (0,2) -- (1,4) -- (2,4) -- (3,2) -- cycle; %

            \draw[thick] (0,0) -- (0,2); %
            \draw[thick] (3,0) -- (3,2); %
            \draw[thick] (0,2) -- (1,4) -- (2,4) -- (3,2); %
            \draw[thick] (0,2) -- (3,2);
            \node at (1.5, 5.5) {3D Scrubs Representation};
        \end{scope}
    \end{tikzpicture}
    \caption{3D representations of tree (left) and scrubs (right). The tree features a cylindrical trunk and crown, while scrubs are modeled as a extruded volume.}
    \label{fig:tree_and_scrubs_visualization}
\end{figure}
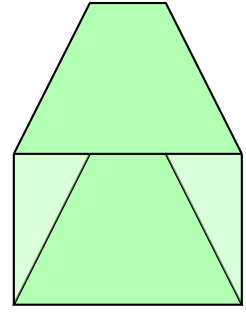

\paragraph{Trees}
Tree geometries are created using cylindrical trunks and spherical crowns:

The trunk is represented by a cylinder with a fixed or parsed radius and height.
The crown is visualized as a sphere placed at the top of the trunk.
Variations in leaf type and height are taken into account to enhance realism, as shown in Figure~\ref{fig:tree_and_scrubs_visualization}.

\paragraph{Result of the geometry extraction}

The methods described above were applied to an OSM map within the bounding box defined by longitude (9.2063390--9.2118430) and latitude (48.4886460-- 48.4918950). This process generated the geometry shown in Figure~\ref{fig}, which illustrates streets, buildings, and porous objects such as trees and shrubs colored green.
This extracted geometry serves as the basis for the simulation setup used in the DT.

\begin{figure}[ht]
    \centering
    \includegraphics[width=1\linewidth]{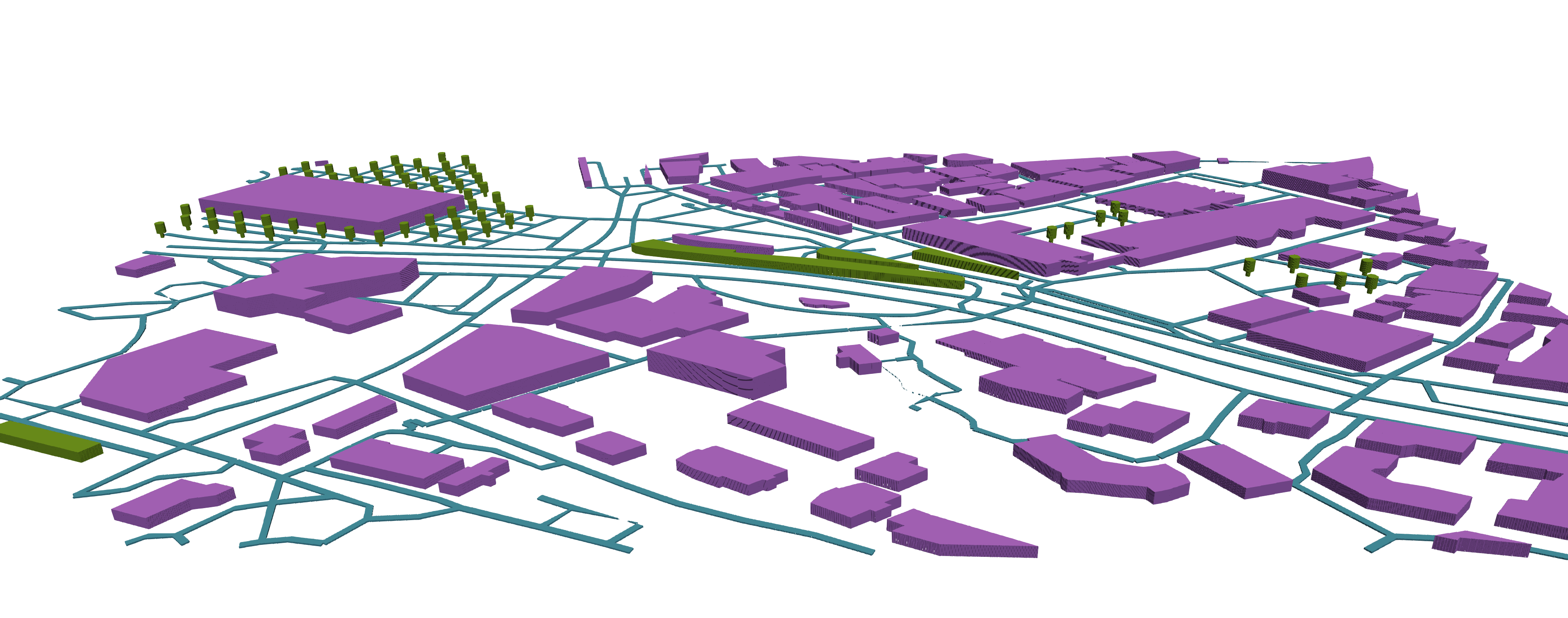}
    \caption{Simulation geometry generated with OSM data. Depicted are buildings in purple, streets in gray and porous objects such as trees and scrub in green.}
    \label{fig:complete_geometry}
\end{figure}

\subsection{Identification of sources}\label{sec:sources}

To address the challenge of identifying sources of pollution, some assumptions must be made. Research indicates that most nitrogen dioxide pollution can be traced back to traffic emissions~\cite{RIJINDERS01}. Although some of the pollution comes from factories, we assume that most is mainly from traffic. With this assumption, we can roughly estimate the number of cars on the street where the measurement station is located.

We first calculate the volume of the street by taking the length $ L $, width $ W $, and height $ H $ of the measuring station, which is approximately 5 meters above the ground. The volume $ V $ is given by
\begin{equation}    
V = L \times W \times H.
\end{equation}

By multiplying the measured concentration $ C $ of nitrogen dioxide with the volume $ V $, we can calculate the total nitrogen dioxide $ \text{TNO}_2 $ on the street:
\begin{equation}    
\text{TNO}_2 = C V.
\end{equation}

To estimate the number of cars $ N $ on the street, we divide the total nitrogen dioxide by the median emissions $ E $ of a car, which is $0.3 \frac{\text{g}}{\text{km}}$:
\begin{equation}    
N = \frac{\text{TNO}_2}{E}.
\end{equation}

In addition to traffic, many other sources can contribute to increased levels of air pollution. With the estimated number of cars, we can calculate the contribution of the particulate matter emitted by the cars. By subtracting that amount from the measured value, we can define the amount contributed by other sources (e.g. buildings).

\subsection{Mathematical modeling of urban areas}\label{sec:modelling}

\begin{figure}[h]
\centering
\includegraphics[width=\textwidth]{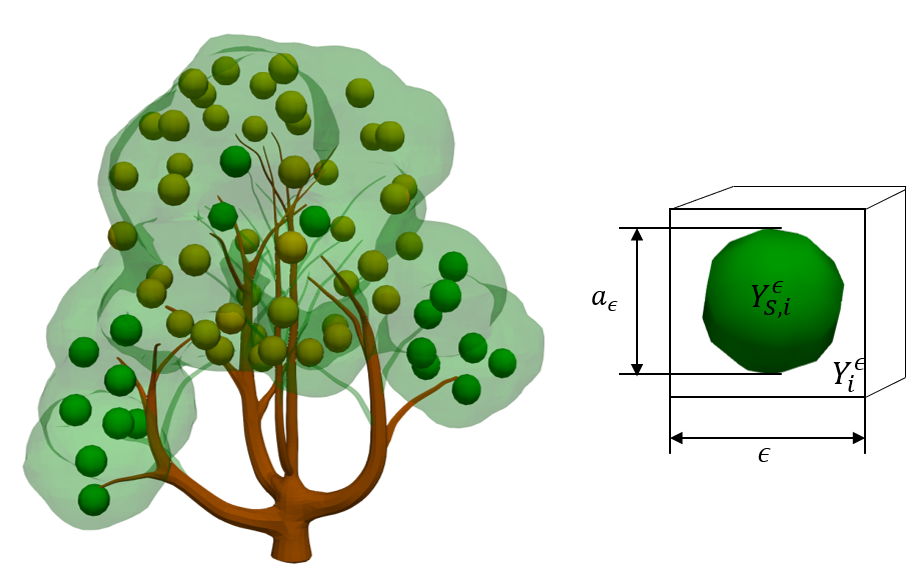}
\caption{Geometric model of a porous medium in three dimensions $(d=3)$. Left: Subvolume of the porous medium, showing the arrangement of spherical obstacles within the matrix. Right: Detailed view of the \(i\)th cell, denoted as $Y_i^\epsilon$, containing a spherical obstacle $Y_{\mathrm{So}, i}^\epsilon$, with radius $a_\epsilon$ . Each cell is a cube with side length $\epsilon$.}
\label{fig}
\end{figure}

Modeling porous objects inside an urban area can be very challenging because of the presence of smaller elements, such as leaves on trees, which are difficult to resolve. To avoid the computational expense of resolving such small objects, we model them and assume an even distribution of solid materials interspersed with empty spaces, as shown in Figure~\ref{fig}. The illustration shows a unified geometric model according to Simonis~\textit{et al.}~\cite{SIMONIS23} of a porous structure, where the unit cells are indicated by $Y^\epsilon_i$ with $\epsilon$ representing the side length. The overall fluid domain $\Omega_\epsilon$ is obtained by removing the solids, i.e. 
\begin{equation}
\Omega_\epsilon = \Omega \setminus \bigcup_{i=1}^{N(\epsilon)} Y^\epsilon_{\mathrm{So},i} , 
\end{equation}
where \(Y_{\mathrm{So},i}^{\epsilon}\) denotes the solid part of a unit cell, and \(N(\epsilon)\) counts their overall number within the porous region \(\Omega\). 

\subsection{Homogenized filtered Brinkman--Navier--Stokes equation}
The Navier--Stokes equations (NSE) for incompressible flow govern the motion of fluids and are fundamental in fluid dynamics. The NSE consist of two primary equations: the mass conservation equation and the momentum equation. To overcome the computational costs of resolving the Kolmogorov scale in turbulent flows, we use a spatial filtering operation that is denoted as \(\bar{\cdot}\,\) below. 
The mass conservation equation ensures that the fluid density remains constant and is expressed as  
\begin{equation}\label{eq:massNSE}
    \bm{\nabla} \cdot \bar{\bm{u}}_\epsilon =0, \quad \text{in } \Omega_\epsilon \times I,
\end{equation}  
where $\bar{\bm{u}}_{\epsilon}$ is the filtered velocity vector of the fluid and \(I\subseteq \mathbb{R}_{>0}\) denotes the time horizon. The momentum equation describes the balance of forces acting on the fluid and is given by  
\begin{equation}
        \frac{\partial \bar{\bm{u}}_\epsilon}{\partial t} + \bar{\bm{u}}_\epsilon \cdot \bm{\nabla} \bar{\bm{u}}_\epsilon = -\frac{\bm{\nabla} \bar{p}_\epsilon}{\rho} + \nu_{\mathrm{mo}} \bm{\nabla}^2 \bar{\bm{u}}_\epsilon + \frac{\bm{F}}{\rho} - \bm{\nabla} \cdot \mathbf{T}, \quad \text{in } \Omega_\epsilon \times I,
\end{equation}  
where $\bar{p}_\epsilon$ is the filtered pressure, $\rho$ is the fluid density, $\nu_{\mathrm{mo}}$ is the molecular kinematic viscosity of the fluid, and $\bm{F}$ is the external force acting on the fluid. The additional term $\bm{\nabla} \cdot \mathbf{T}$ represents the effects of filtering approximated, for example, as small-scale turbulence, modeled using a subgrid-scale stress tensor in large eddy simulation (LES).  

Fluid flow through porous media is strongly influenced by the porous structure, which adds additional resistance to the motion of the fluid. To account for this, the NSE on an obstacle level (resolving the porous media) are homogenized with a critical obstacle size (cf.\ Simonis~\textit{et al.}~\cite{SIMONIS23}). 
The homogenization limit in this case results in incorporating the Brinkman-term while also applying a filtering operation to capture large-scale turbulence effects using LES. This yields the filtered Brinkman--Navier--Stokes equations (FBNSE) 
\begin{align}
\begin{cases}
    \bm{\nabla} \cdot \bar{\bm{u}}  =0, & \quad \text{in } \Omega \times I, \\
\frac{\partial \bar{\bm{u}}}{\partial t} + \bar{\bm{u}} \cdot \bm{\nabla} \bar{\bm{u}} = -\frac{\bm{\nabla} \bar{p}}{\rho} + \nu_{\mathrm{mo}} \bm{\nabla}^2 \bar{\bm{u}} + \frac{\nu_{\mathrm{mo}}}{K} \bar{\bm{u}} - \bm{\nabla} \cdot \mathbf{T}_{\mathrm{sgs}}, & \quad \text{in } \Omega \times I, \label{eq:fbnse}
\end{cases}
\end{align}  
where $K>0$ is the permeability coefficient of the porous medium. The term $\frac{\nu_{\mathrm{mo}}}{K} \bar{\bm{u}}$ represents the additional resistance due to the porous structure, which is proportional to the viscosity of the fluid and inversely proportional to the permeability. 
In \eqref{eq:fbnse}, we approximate \(\mathbf{T} \approx \mathbf{T}_{\mathrm{sgs}}\) such that the term $\bm{\nabla} \cdot \mathbf{T}_{\mathrm{sgs}}$ accounts for the subgrid-scale turbulence modeled by the Smagorinsky LES approach  
\begin{align}
    \mathbf{T}_{\mathrm{sgs}} & = 2 \nu_\mathrm{turb} \bar{\mathbf{S}}, \label{eq:sgsStress}\\
    \nu_\mathrm{turb} & = \left(C_{\mathrm{S}} \triangle_{\bm{x}}\right)^2 \left|\bar{\mathbf{S}}\right|, \label{eq:turbVisc}
\end{align}  
where $C_{\mathrm{S}}>0$ is the Smagorinsky constant, $\triangle_{\bm{x}}$ is the filter width, and $\bar{\mathbf{S}}$ is the filtered strain rate tensor:  
\begin{equation}
    \bar{S}_{\alpha\beta} = \frac{1}{2} \left( \frac{\partial \bar{u}_{\alpha}}{\partial x_\beta} + \frac{\partial \bar{u}_{\beta}}{\partial x_\alpha} \right).
\end{equation} 
In particular, in the simulations conducted in this work, we make use of a composition of pure fluid regions and homogenized porous media regions in the overall model such that \(K\) in the Brinkman-term becomes space-dependent. 

However, it is essential to account for not only advection effects but also the influence that porous media have on diffusion. 
The advection--diffusion equation (ADE) for porous media based on the formulation according to Lasaga~\textit{et al.}~\cite{LASAGA14} is given as
\begin{align}
    \frac{\partial (\Phi C)}{\partial t} + \bm{\nabla} \cdot (\Phi C \bar{\bm{u}}) = D \bm{\nabla} \cdot (\Phi \bm{\nabla} C), \quad \text{in } \Omega \times I, \label{eq:porousADE}
\end{align}
where $C$ is the concentration, $D$ is the diffusion coefficient, and $\Phi$ represents the local porosity. 
Together with the initial and boundary conditions considered in Section~\ref{subsubsec:setup}, \eqref{eq:fbnse} and \eqref{eq:porousADE} form the mathematical model considered in this work.  

\subsection{Numerical methods and discretization}
\subsubsection{The homogenized lattice Boltzmann method for fluid flows}
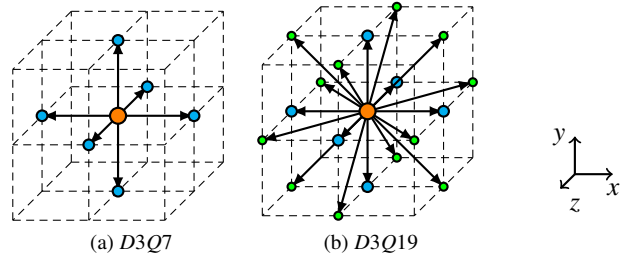
\begin{figure}[ht!]
\centering
\subfloat[\(D3Q7\)]{
\centering
\begin{tikzpicture}[>=latex]
    \draw[densely dashed] (-2,0,0) -- (0,0,0);
    \draw[densely dashed] (-2,0,0) -- (-2,2,0);
    \draw[densely dashed] (-2,0,0) -- (-2,0,2);
    \draw[densely dashed] (-2,0,2) -- (0,0,2);
    \draw[densely dashed] (-2,2,0) -- (0,2,0);
    \draw[densely dashed] (-2,0,2) -- (-2,2,2);
    \draw[densely dashed] (-2,2,0) -- (-2,2,2);
    \draw[densely dashed] (-2,2,2) -- (0,2,2);
    \draw[densely dashed] (0,0,0) -- (0,0,2);
    \draw[densely dashed] (0,0,0) -- (0,2,0);
    \draw[densely dashed] (0,0,2) -- (0,2,2);
    \draw[densely dashed] (0,2,2) -- (0,2,0);
 	\draw[densely dashed] (-2,0,1) -- (0,0,1); 
	\draw[densely dashed] (-2,0,1) -- (-2,2,1); 
	\draw[densely dashed] (-2,2,1) -- (0,2,1); 
	\draw[densely dashed] (0,2,1) -- (0,0,1); 
	\draw[densely dashed] (-1,0,0) -- (-1,0,2); 
	\draw[densely dashed] (-1,0,2) -- (-1,2,2); 
	\draw[densely dashed] (-1,2,2) -- (-1,2,0); 
	\draw[densely dashed] (-1,2,0) -- (-1,0,0); 
	\draw[densely dashed] (-2,1,0) -- (0,1,0); 
	\draw[densely dashed] (0,1,0) -- (0,1,2); 
	\draw[densely dashed] (0,1,2) -- (-2,1,2); 
	\draw[densely dashed] (-2,1,2) -- (-2,1,0); 
	\draw[->,thick](-1,1,1) -- (-1,1,0);
    \draw[->,thick](-1,1,1) -- (-1,2,1);
    \draw[->,thick](-1,1,1) -- (-1,1,2);
    \draw[->,thick](-1,1,1) -- (0,1,1);
    \draw[->,thick](-1,1,1) -- (-2,1,1);
    \draw[->,thick](-1,1,1) -- (-1,0,1);
    \draw[thick,fill=orange](-1,1,1) circle(3pt);
	\draw[thick,fill=cyan](-1,1,0) circle(2pt);     
	\draw[thick,fill=cyan](-1,2,1) circle(2pt);
    \draw[thick,fill=cyan](-1,1,2) circle(2pt);
    \draw[thick,fill=cyan](0,1,1) circle(2pt);
    \draw[thick,fill=cyan](-2,1,1) circle(2pt);
    \draw[thick,fill=cyan](-1,0,1) circle(2pt);
\end{tikzpicture}\hspace{1em}
}
\subfloat[\(D3Q19\)]{\begin{tikzpicture}[>=latex]
 \centering
    \draw[densely dashed] (-2,0,0) -- (0,0,0);
    \draw[densely dashed] (-2,0,0) -- (-2,2,0);
    \draw[densely dashed] (-2,0,0) -- (-2,0,2);
    \draw[densely dashed] (-2,0,2) -- (0,0,2);
    \draw[densely dashed] (-2,2,0) -- (0,2,0);
    \draw[densely dashed] (-2,0,2) -- (-2,2,2);
    \draw[densely dashed] (-2,2,0) -- (-2,2,2);
    \draw[densely dashed] (-2,2,2) -- (0,2,2);
    \draw[densely dashed] (0,0,0) -- (0,0,2);
    \draw[densely dashed] (0,0,0) -- (0,2,0);
    \draw[densely dashed] (0,0,2) -- (0,2,2);
    \draw[densely dashed] (0,2,2) -- (0,2,0);
 	\draw[densely dashed] (-2,0,1) -- (0,0,1); 
	\draw[densely dashed] (-2,0,1) -- (-2,2,1); 
	\draw[densely dashed] (-2,2,1) -- (0,2,1); 
	\draw[densely dashed] (0,2,1) -- (0,0,1); 
	\draw[densely dashed] (-1,0,0) -- (-1,0,2); 
	\draw[densely dashed] (-1,0,2) -- (-1,2,2); 
	\draw[densely dashed] (-1,2,2) -- (-1,2,0); 
	\draw[densely dashed] (-1,2,0) -- (-1,0,0); 
	\draw[densely dashed] (-2,1,0) -- (0,1,0); 
	\draw[densely dashed] (0,1,0) -- (0,1,2); 
	\draw[densely dashed] (0,1,2) -- (-2,1,2); 
	\draw[densely dashed] (-2,1,2) -- (-2,1,0);   
    \draw[->,thick](-1,1,1) -- (-1,1,0);
    \draw[->,thick](-1,1,1) -- (-2,1,0);
    \draw[->,thick](-1,1,1) -- (-1,2,0);
    \draw[->,thick](-1,1,1) -- (-1,0,0);
    \draw[->,thick](-1,1,1) -- (0,1,0);
    \draw[thick,fill=cyan](-1,1,0) circle(2pt);     
	\draw[thick,fill=green](-2,1,0) circle(1.5pt);
  	\draw[thick,fill=green](-1,2,0) circle(1.5pt);
  	\draw[thick,fill=green](-1,0,0) circle(1.5pt);
	\draw[thick,fill=green](0,1,0) circle(1.5pt);
    \draw[->,thick](-1,1,1) -- (-1,2,1);
    \draw[->,thick](-1,1,1) -- (0,1,1);
    \draw[->,thick](-1,1,1) -- (-2,1,1);
    \draw[->,thick](-1,1,1) -- (-1,0,1);
    \draw[->,thick](-1,1,1) -- (-2,0,1);
    \draw[->,thick](-1,1,1) -- (0,0,1);
    \draw[->,thick](-1,1,1) -- (-2,2,1);
    \draw[->,thick](-1,1,1) -- (0,2,1);
	\draw[thick,fill=cyan](-1,2,1) circle(2pt);
    \draw[thick,fill=cyan](0,1,1) circle(2pt);
    \draw[thick,fill=cyan](-1,0,1) circle(2pt);
    \draw[thick,fill=cyan](-2,1,1) circle(2pt);
    \draw[thick,fill=green](-2,0,1) circle(1.5pt);     
  	\draw[thick,fill=green](0,0,1) circle(1.5pt);
    \draw[thick,fill=green](-2,2,1) circle(1.5pt);     
  	\draw[thick,fill=green](0,2,1) circle(1.5pt);
	\draw[->,thick](-1,1,1) -- (-1,1,2);
    \draw[->,thick](-1,1,1) -- (-1,0,2);
	\draw[->,thick](-1,1,1) -- (-2,1,2);
	\draw[->,thick](-1,1,1) -- (0,1,2);
	\draw[->,thick](-1,1,1) -- (-1,2,2);
    \draw[thick,fill=cyan](-1,1,2) circle(2pt);
	\draw[thick,fill=green](-1,0,2) circle(1.5pt);
	\draw[thick,fill=green](-2,1,2) circle(1.5pt);
	\draw[thick,fill=green](0,1,2) circle(1.5pt);
	\draw[thick,fill=green](-1,2,2) circle(1.5pt);
    \draw[thick,fill=orange](-1,1,1) circle(3pt);
  \end{tikzpicture}
  }
\hspace{2em}
\begin{tikzpicture}
	\draw[->,thick] (0,0,0) -- (.5,0,0);
    \node[anchor=north] at (.5,0,0) {$x$};
    \draw[->,thick] (0,0,0) -- (0,.5,0);
    \node[anchor=east] at (0,.5,0) {$y$};
	\draw[->,thick] (0,0,0) -- (0,0,.5);
	\node[anchor=north west] at (0,0,.5) {$z$};
\end{tikzpicture}
  \caption{A schematic illustration of the discrete velocity sets (a) \(D3Q7\) and (b) \(D3Q19\). 
  Coloring refers to energy shells: orange, cyan, green denote zeroth, first, second order, respectively. 
  Figure from \cite{simonis2023pde}.}
\label{fig:lattice}
\end{figure}

Due to the high computational costs required for our DT framework, we use the HLBM to discretize the FBNSE \eqref{eq:fbnse} on a space-time grid, using the \(D3Q19\) velocity stencil illustrated in Figure~\ref{fig:lattice}. The HLBM inherits the classical advantages of LBM in terms of high parallelizability and efficient scaling with increasing hardware capabilities \cite{KUMEERLAENDER22}. 
In this work, we extend the HLBM initially proposed by Krause~\textit{et al.}~\cite{KRAUSE20171} with the hybrid third-order recursive regularized collision model proposed by Jacob~\textit{et al.}~\cite{JACOB18}. 
First, we summarize the resulting space-time evolution equation and, afterwards, specify the individual features. 
Note that, for compactness of notation, we abandon the \(\bar{\cdot}\,\)-notation of the filtered quantities below. 

The filtered and homogenized lattice Boltzmann equation approximating the FBNSE \eqref{eq:fbnse} is given by  
\begin{equation}\label{eq:hlbm}
    f_{i} (\bm{x}+\bm{c}_i \triangle t, t+\triangle t) 
    = 
    f_{i}^{\mathrm{h},\mathrm{eq}} (\bm{x}, t) + \left( 1 - \frac{1}{\tau_{\mathrm{eff}}(\bm{x},t)} \right)\tilde{f}_{i}^{(1)}(\bm{x}, t), \quad \text{in } \Omega_{\triangle x} \times I_{\triangle t}, 
\end{equation}
where \( f_i: \Omega_{\triangle x} \times I_{\triangle t} \rightarrow \mathbb{R}_{\geq 0} \) represents the filtered distribution functions, \(\bm{c}_i\) denote the \(q\) mesoscopic velocities in \(D3Q19\), where \(i=0, 1, \dots, q-1\), \( \Omega_{\triangle x} \subset \Omega \subseteq \mathbb{R}^3\) is the discretized position space domain with voxel size \( \triangle x \), and \(I_{\triangle t} \subset I \subseteq \mathbb{R}_{\geq 0}\) is the discrete time horizon with timestep size \(\triangle t\). 
Moreover, \( \tau_{\mathrm{eff}} \) is the space-time adaptive, effective relaxation time of the naive Smagorinsky Bhatnagar--Groos--Krook (BGK) model, given by  
\begin{equation}\label{eq:tauEff}
    \tau_\mathrm{eff}(\bm{x},t) = \frac{\nu_\mathrm{eff}(\bm{x},t)}{c_{\mathrm{s}}^2} \frac{\triangle t}{\triangle x^2} + \frac{1}{2},
\end{equation}
with \( \nu_\mathrm{eff} = \nu_{\mathrm{mo}} + \nu_\mathrm{turb} \) representing the combined molecular and turbulent viscosity \eqref{eq:turbVisc} and \( c_{\mathrm{s}} \) is the lattice speed of sound. 
The regularization in \eqref{eq:hlbm} is based on 
the non-equilibrium function $f_{i}^{(1)} = f_{i} - f_{i}^{(0)}$ that is expanded as
\begin{equation}
  f_{i}^{(1)}(\bm{x},t) = \omega_i \sum_{n=0}^{N=3} \frac{1}{c_{\mathrm{s}}^{2n} n!} \mathbf{H}_i^{(n)} : \mathbf{a}_1^{(n)}(\bm{x},t) ,
\end{equation}
where \(\omega_i\) are the lattice weights and we denote by \( \mathbf{H}_i^{(n)} \) the \(n\)th order Hermite polynomial with the \(i\)th discrete velocity \(\bm{c}_{i}\) as an argument. 
The Hermite coefficient for the non-equilibrium is defined as
\begin{equation}
    \mathbf{a}_1^{(n)}(\bm{x},t)=\sum_{i=0}^{q-1}\mathbf{H}_i^{(n)}f_i^{(1)}(\bm{x},t),
\end{equation}
Note that according to \cite{JACOB18}, we use Hermite polynomials that have correct orthogonality for \(D3Q19\) only. 
Then we add the hybridization of rate of strain via 
\begin{align}
    \tilde{f}_{i}^{(1)}(\bm{x},t) = f_{i}^{(1)} (\bm{x},t)\sigma  - (1-\sigma) \frac{\rho \tau}{c_{\mathrm{s}}^{2}} \mathbf{H}_{i}^{(2)} : \mathbf{S}^{\mathrm{FD}} (\bm{x},t), 
\end{align}
where \(0\leq \sigma \leq 1\) and 
\begin{align}
    S^{\mathrm{FD}}_{\alpha\beta} \approx S_{\alpha\beta} + \frac{1}{2}\left[\frac{1}{6} \triangle x^{2} \left( \frac{\partial^{3} u_{\beta}}{\partial x_{\alpha}^{3}} +  \frac{\partial^{3} u_{\alpha}}{\partial x_{\beta}^{3}} \right)\right]
\end{align} 
is the finite difference (FD) strain rate tensor. 
Notably, \(\sigma = 1 \) switches off the FD contribution and reduces the model to a recursively regularized, homogenized Smagorinsky BGK collision. 
Further, extending the method to homogenized fluid flow, we define the homogenized equilibrium Hermite coefficients \(\widehat{\mathbf{a}}_{0}^{(n)} = \mathbf{a}_{0}^{(n-1)} \widehat{\bm{u}}\) with \(\widehat{\mathbf{a}}_{0}^{(0)} = \rho\). 
Then, the homogenized equilibrium distribution function is defined as
\begin{equation}
  f_i^{\mathrm{h},\mathrm{eq}}(\bm{x},t) = \omega_i \left(\rho(\bm{x},t) + \frac{\bm{c}_i \cdot (\rho\, \mathbf{\widehat{u}}(\bm{x},t))}{c_{\mathrm{s}}^2}
    + \frac{\mathbf{H}_i^{(2)} : \widehat{\mathbf{a}}_0^{(2)}(\bm{x},t)}{2 c_{\mathrm{s}}^4}
    + \frac{\mathbf{H}_i^{(3)} : \widehat{\mathbf{a}}_0^{(3)}(\bm{x},t)}{2 c_{\mathrm{s}}^6}
  \right),
\end{equation}
where \( \rho \) is the zeroth order density moment and \(\widehat{\bm{u}}\) is the homogenized macroscopic velocity field that incorporates the permeability effects in \eqref{eq:fbnse} from the partial homogenization of the domain. 
In the general case (moving solid objects \cite{KRAUSE20171}) the homogenized velocity is defined as a convex combination of the fluid velocity moment \( \bm{u} \) and the solid object velocity \( \bm{u}^{\mathrm{B}}\), given by  
\begin{equation}\label{eq:convexVelocity}
    \widehat{\bm{u}}(\bm{x},t) = (1 - d(\bm{x},t)) \bm{u}(\bm{x},t) + d(\bm{x},t) \bm{u}^{\mathrm{B}}(\bm{x},t), 
\end{equation}
where \(d\) is the so-called lattice porosity. 
In the present case (rigid obstacles, i.e.\ \( \bm{u}^{\mathrm{B}} = 0 \)), \eqref{eq:convexVelocity} reduces to  
\begin{equation}
    \widehat{\bm{u}}(\bm{x},t) = (1 - d(\bm{x},t)) \bm{u}(\bm{x},t), 
\end{equation}
where  
\begin{equation}
    d(\bm{x},t) = 1 - \frac{\triangle x^2 \nu \tau_{\mathrm{mo}}}{K(\bm{x},t)}, 
    \label{eq:lattice_porosity}
\end{equation}
the permeability given in \eqref{eq:fbnse} is identified with \(K\) and \(\tau_{\mathrm{mo}}\) is the molecular relaxation time, defined similarly as in \eqref{eq:tauEff} but with \(\nu_{\mathrm{mo}}\). 

In \cite{SIMONIS23}, a standalone derivation and analysis of the continuous homogenized kinetic model with a BGK collision for fluid flow in porous media is proposed. 
For unfiltered fields, a formal estimate for the order of approximation is derived in \cite{simonis2023pde}. 
In general, the fluid velocity moment \(\bm{u}\) from \eqref{eq:hlbm} is expected to provide a second-order approximation in space of a filtered velocity solution of \eqref{eq:fbnse}.

\subsection{The homogenized lattice Boltzmann method for advection--diffusion processes}

So far, HLBM \eqref{eq:hlbm} recovers the flow in and around porous media modeled by \eqref{eq:fbnse}. 
When coupled with a discretization of the ADE \eqref{eq:porousADE}, the overall numerical method is able to reflect the unique behavior of the particles within a porous environment. 
To incorporate these effects into the HLBM, a second lattice Boltzmann equation is coupled to \eqref{eq:hlbm} that includes a correction term $\mathcal{R}$. 
This term is derived below and accounts for temporal and spatial variations in porosity, enabling an accurate reflection of the local concentration within the porous media.

The correction term is derived by expanding each component of \eqref{eq:porousADE} separately and isolating the temporal and spatial variations of $\Phi$.
The first term of \eqref{eq:porousADE} is expanded using the time derivative as 
\begin{equation} 
\frac{\partial (\Phi C)}{\partial t} = \Phi \frac{\partial C}{\partial t} + C \frac{\partial \Phi}{\partial t}. 
\label{eq:deria1}
\end{equation}
Similarly, the advective term is expanded as 
\begin{equation} \bm{\nabla} \cdot (\Phi C \bm{u}) = \Phi \bm{u} \cdot \bm{\nabla} C + C \bm{u} \cdot \bm{\nabla} \Phi + C \Phi (\bm{\nabla} \cdot \bm{u}), 
\label{eq:deria2}
\end{equation}
and the diffusive term expands to 
\begin{equation} D \bm{\nabla} \cdot (\Phi \bm{\nabla} C) = D (\bm{\nabla} \Phi) \cdot \bm{\nabla} C + D \Phi \bm{\nabla}^2 C. 
\label{eq:deria3} 
\end{equation}
Combining the temporal and spatial variations from \eqref{eq:deria1} to \eqref{eq:deria3}, we obtain the correction term 
\begin{equation}
    \mathcal{R} 
    =
    - C \frac{\partial_t \Phi}{\partial t} \frac{1}{\Phi} - C \bm{u}\left(1+D \bm{\nabla} C \right)\frac{\bm{\nabla} \Phi}{\Phi} .
    \label{eq:correctionTermCont}
\end{equation}
With a finite difference approximation of \eqref{eq:correctionTermCont} we get
\begin{align}
\mathcal{R} (\bm{x}_{\bm{j}}, t_{n} ) 
    \approx \; & \mathcal{R}_{\triangle x, \triangle t} (\bm{x}_{\bm{j}}, t_{n} ) \\
    =  &- C_{\bm{j}}^n \frac{\Phi_{\bm{j}}^n - \Phi_{\bm{j}}^{n-1}}{\Phi_{\bm{j}}^n \triangle t} \nonumber\\
    &- C_{\bm{j}}^n \left( u_x \frac{\Phi_{\bm{j} + (1,0,0)}^n - \Phi_{\bm{j} - (1,0,0)}^n}{2 \Phi_{\bm{j}}^n \triangle x} + 
    u_y \frac{\Phi_{\bm{j} + (0,1,0)}^n - \Phi_{\bm{j} - (0,1,0)}^n}{2 \Phi_{\bm{j}}^n \triangle x} + 
    u_z \frac{\Phi_{\bm{j} + (0,0,1)}^n - \Phi_{\bm{j} - (0,0,1)}^n}{2 \Phi_{\bm{j}}^n \triangle x} \right) \nonumber\\
    &+ D \left( \frac{\left(C_{\bm{j} + (1,0,0)}^n - C_{\bm{j} - (1,0,0)}^n \right) \left(\Phi_{\bm{j} + (1,0,0)}^n - \Phi_{\bm{j} - (1,0,0)}^n \right)}{4 \Phi_{\bm{j}}^n \triangle x^2} \right . \nonumber\\ 
    &\qquad~\left . + \frac{\left( C_{\bm{j} + (0,1,0)}^n - C_{\bm{j} - (0,1,0)}^n \right) \left(\Phi_{\bm{j} + (0,1,0)}^n - \Phi_{\bm{j} - (0,1,0)}^n \right)}{4 \Phi_{\bm{j}}^n \triangle x^2} \right . \nonumber\\
    &\qquad~\left .+\frac{\left( C_{\bm{j} + (0,0,1)}^n - C_{\bm{j} - (0,0,1)}^n \right) \left( \Phi_{\bm{j} + (0,0,1)}^n - \Phi_{\bm{j} - (0,0,1)}^n \right)}{4 \Phi_{\bm{j}}^n \triangle x^2} \right)
     \label{eq:finiteApprox}
\end{align}
for \(\cdot\,(\bm{x}_{\bm{j}}, t_{n}) = \cdot\,_{\bm{j}}^{n}\) at grid nodes \(\bm{x}_{\bm{j}} \in \Omega_{\triangle x}\) and time steps \(t_{n} \in I_{\triangle t}\), which is second order in space and first order in time. 
Asymmetric stencils of the same order are used at the domain boundaries. 
The LBM discretization of the advection--diffusion process, including a modified collision operator to account for spatial porosity variations through homogenization, is expressed as
\begin{equation} 
    g_i(\bm{x} + \bm{c}_i \triangle t, t + \triangle t) - g_i(\bm{x}, t) 
    = 
    \frac{1}{\tau_{D}}\left( g_i^{\mathrm{eq}}(\bm{x},t)-g_i(\bm{x},t)\right) + J^{\mathrm{por}}(\bm{x},t), \quad \text{in } \Omega_{\triangle x} \times I_{\triangle t}, \label{eq:corrAde} 
\end{equation}
where \(\tau_{D}\) is connected to the diffusivity \(D\) similarly as in \eqref{eq:turbVisc}. 
The equilibrium distribution function $g_i^{\mathrm{eq}}$, representing the equilibrium concentration distribution, is given by
\begin{equation} \label{eq:adeEQ}
    g_i^{\mathrm{eq}}(\bm{x}, t) = w_i C (\bm{x},t)\left( 1 + \frac{\bm{c}_i \cdot \bm{u}(\bm{x},t)}{c_{\mathrm{s}}^2} \right), 
\end{equation}
where \(\bm{u}\) is the velocity moment from \eqref{eq:hlbm} and the concentration $C$ is calculated as the zeroth moment of the populations $g_i$ with an additional term derived from the second-order finite difference approximation of \eqref{eq:correctionTermCont} named \(\mathcal{R}_{\triangle x, \triangle t}\) and defined in \eqref{eq:finiteApprox}, i.e. 
\begin{equation} 
    C (\bm{x},t) = \sum_i g_i (\bm{x},t)+ \frac{1}{2} \mathcal{R}_{\triangle x, \triangle t}(\bm{x},t) .  
\end{equation}
The additional collision term, $J^{\mathrm{por}}$ in \eqref{eq:corrAde}, is introduced to adjust the transport dynamics in response to spatial porosity variations and reads 
\begin{equation} 
    J^{\mathrm{por}} (\bm{x}, t ) 
    = 
    \left( 1 - \frac{1}{2 \tau_{D}} \right) w_i \mathcal{R}_{\triangle x, \triangle t} (\bm{x}, t). 
\label{eq:correctionTerm}
\end{equation}
The term \eqref{eq:correctionTerm} modifies the overall collision operator to ensure that the equilibrium distribution and the transport properties adapt to local porosity changes. 
The porosity \(\Phi\) is incorporated in \eqref{eq:corrAde} by using the lattice porosity \(d(\bm{x}, t)\) from \eqref{eq:lattice_porosity} in the correction term \(\mathcal{R}_{\triangle x, \triangle t}\). 
Due to the linearity of \eqref{eq:adeEQ}, we use the \(D3Q7\) velocity set (see Figure~\ref{fig:lattice}) to reduce computational effort. 
The second order in space of approximations by LBMs for advection--diffusion equations has been proven among others by Simonis~\textit{et al.}~\cite{simonis2023pde,simonis2020relaxation,simonis2022constructing}.

\section{Validation}\label{sec:validation}
In this section, we present the validation of our numerical approach by comparing simulation results against analytical solutions and experimental data. First, we validate the proposed double-distribution HLBM with a correction term using an analytical benchmark, ensuring the accuracy of the method in solving coupled FBNSE and ADE systems in porous media. Second, we assess the simulation results against wind channel experiments to confirm the applicability of the model for real-world urban flow scenarios.

\subsection{Validation of HLBM with correction term} 

\begin{figure}[ht]
    \centering
\begin{tikzpicture}
\begin{axis}[
    width=0.7\textwidth, %
    xlabel={Resolution},
    ylabel={Error},
    xmode=log, %
    ymode=log, %
    log basis x=10, %
    log basis y=10, %
    legend style={
        at={(0,0)}, %
        anchor=south west, %
        legend columns=1, %
        /tikz/every even column/.append style={column sep=0.5cm} %
    },
    legend cell align={left}, %
    grid=both, %
    minor grid style={dotted}, %
    major grid style={solid, gray!30} %
]

\addplot[color=blue, mark=square*, line width=1pt] 
    table [x={Resolution}, y={L1_abs_Error}, col sep=comma] 
    {error_porousADE.csv};
\addlegendentry{$L^1$}

\addplot[color=green, mark=o, line width=1pt] 
    table [x={Resolution}, y={L2_abs_Error}, col sep=comma] 
    {error_porousADE.csv};
\addlegendentry{$L^2$}

\addplot[color=red, mark=triangle*, line width=1pt] 
    table [x={Resolution}, y={Linf_abs_Error}, col sep=comma] 
    {error_porousADE.csv};
\addlegendentry{$L^\infty$}

\end{axis}
\end{tikzpicture}
\caption{Experimental order of convergence of the error norms $L^1$, $L^2$ and $L^\infty$ to an analytical solution. It shows that the error gets smaller with higher resolution.}
\label{fig:eoc}
\end{figure}
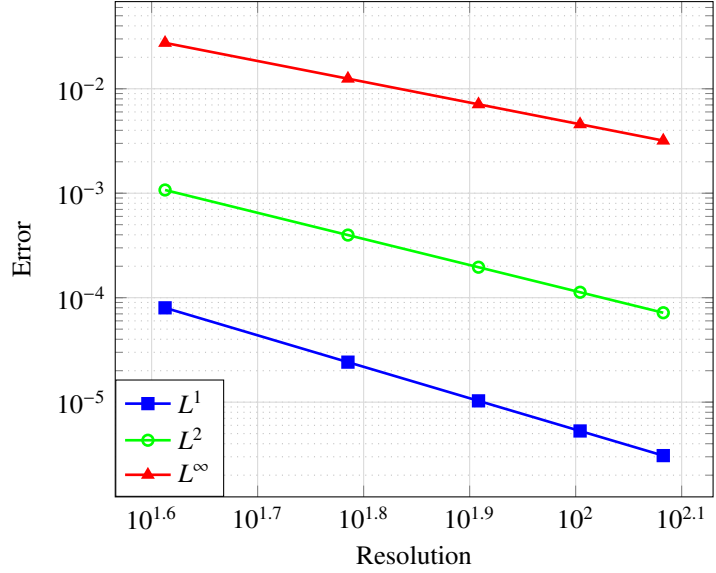

To validate the effectiveness of the HLBM (\eqref{eq:hlbm} and \eqref{eq:corrAde}) with the correction term \eqref{eq:finiteApprox} introduced to approximate the FBNSE \eqref{eq:fbnse} coupled to the ADE \eqref{eq:porousADE} in porous media, the numerical results are compared with an analytical solution. The analytical solution serves as a benchmark and is given by the following equations:
\begin{align}
    \Phi(\bm{x}, t) & = 0.5 + 0.4 \sin(2\pi( x - u_{\mathrm{p}} t)), \\
    C(\bm{x}, t) & =\frac{1}{\Phi(\bm{x}, t)}, \\
    \bm{u}(\bm{x}, t ) & = D \Phi(\bm{x}, t)\bm{\nabla} C(\bm{x}, t).
\end{align}
The analytical solution represents a sinusoidal scalar field $\Phi(\bm{x}, t)$ that propagates in the $x$-direction with a propagation speed $u_{\mathrm{p}}$. This form is commonly used in theoretical studies, as it satisfies the ADE under idealized conditions, incorporating both advection (through $u_{\mathrm{p}}$) and diffusion (through $D$). 
To evaluate the performance of the HLBM approach, the error was analyzed in terms of different norms ($L^1$, $L^2$ and $L^\infty$) across a range of resolutions. The experimental order of convergence (EOC) in Figure~\ref{fig:eoc} illustrates the convergence of the error to the analytical solution as the resolution improves. The consistent decrease in error for all norms confirms the stability and accuracy of the method. Specifically, the second-order observed EOC matches the theoretical expectations, demonstrating that the correction term successfully enhances the numerical approach.

The results validate the applicability of the HLBM with the correction term for modeling ADEs in porous media. This is particularly relevant for complex porous structures, such as urban environments, where accurate simulations of pollutant transport are essential. The observed convergence trends highlight the robustness of the method.

\subsection{Validation with experimental data from a wind channel}
In order to validate the simulation part of the DT with the HLBM approach we use the work of Gromke~\textit{et al.}~\cite{GROMKE08} where they validate the CFD results from the software FLUENT with measurements from a wind channel. The geometric setup is made up of two solid square blocks, which represent buildings with the between being a street canyon with a porous zone representing a tree in the middle as shown in Figure~\ref{fig:geometrical_setup}. 

\subsubsection{Setup}\label{subsubsec:setup}
\begin{figure}[h]
\centering
\subfloat{
    \includegraphics[width=0.75\textwidth]{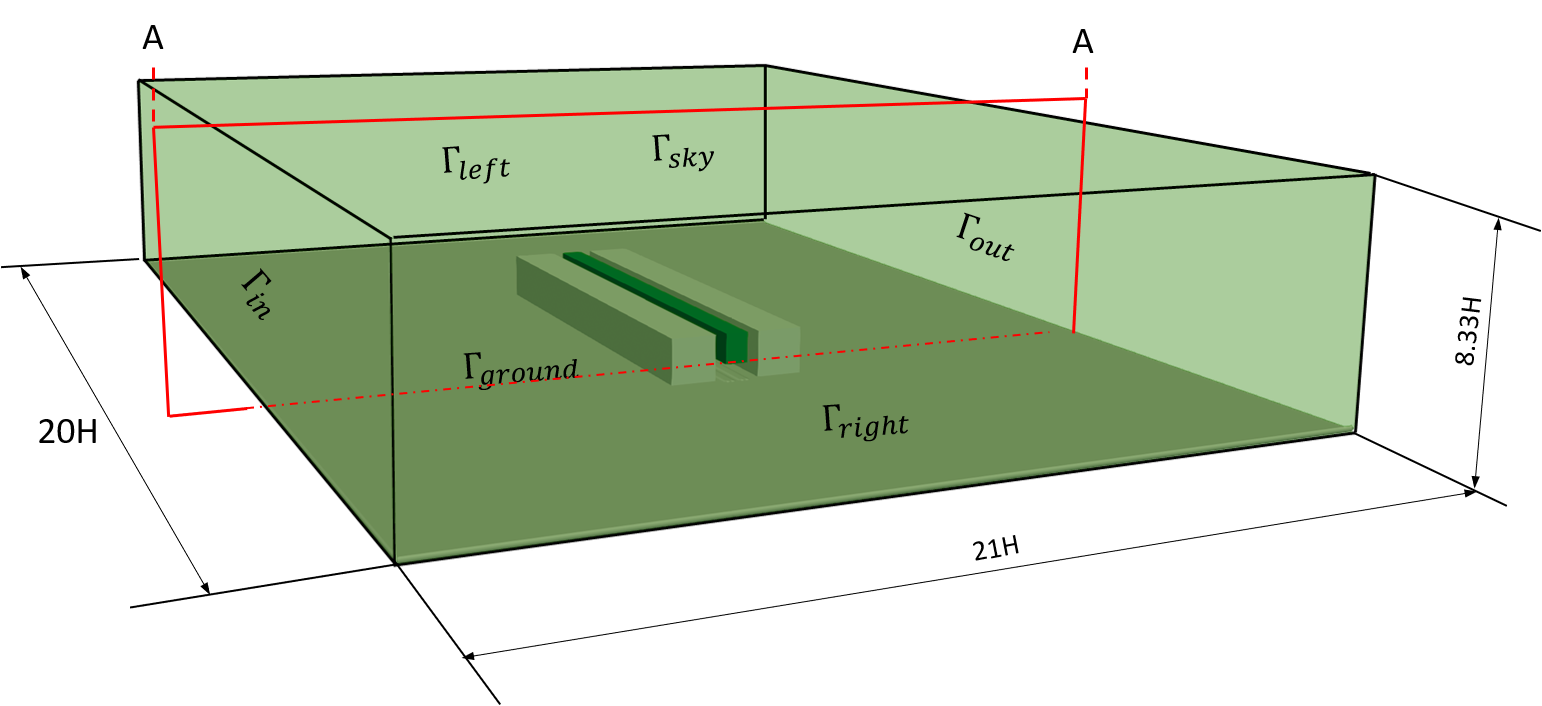}
}
\hfill
\subfloat{
    \includegraphics[width=0.75\textwidth]{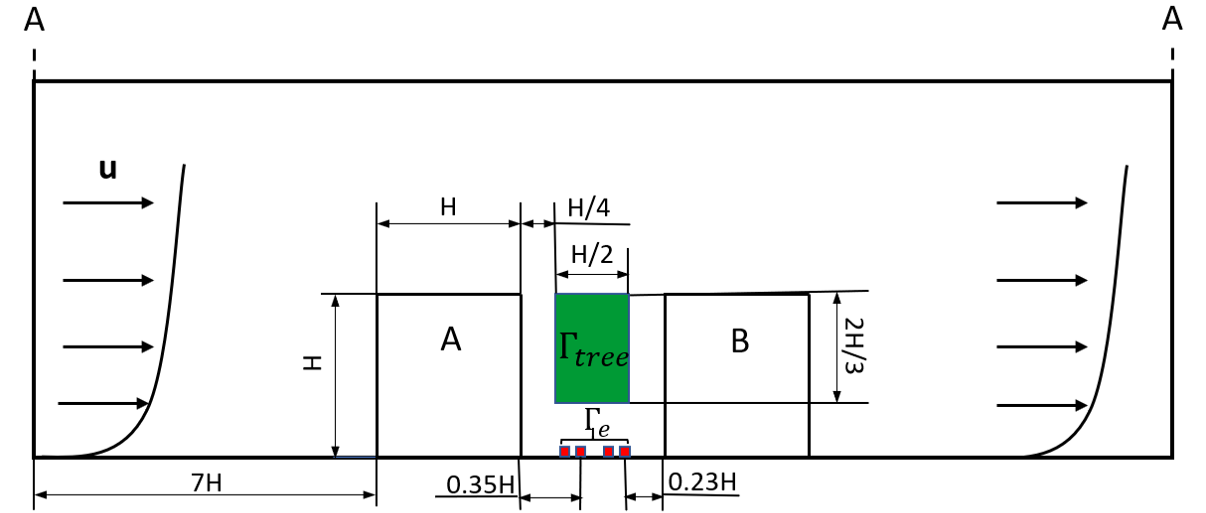}
}
    \caption{Simulation setup for airflow and pollutant distribution in an urban environment. The 3D view (top) shows the boundary conditions ($\Gamma$) and placement of urban structures within the simulation domain. The 2D cross-sectional view (bottom) illustrates the wind profile and flow around a street canyon (A and B), a tree ($\Gamma_\mathrm{tree}$), and four line emitters ($\Gamma_\mathrm{e}$) within the domain.}
    \label{fig:geometrical_setup}
\end{figure}

Two setups were used for the simulations, both following the geometrical setup of the experiment by Gromke~\textit{et al.}~\cite{GROMKE08} and the simulation setup from~\cite{MERLIER18}, as shown in Figure~\ref{fig:geometrical_setup}. 
The complete boundary configuration is described as
\begin{align}
    \Gamma &= \partial\Omega = \Gamma_\mathrm{in} \cup \Gamma_\mathrm{out} \cup \Gamma_\mathrm{ground} \cup \Gamma_\mathrm{wall} \cup \Gamma_\mathrm{sky} \cup \Gamma_\mathrm{left} \cup \Gamma_\mathrm{right} \cup \Gamma_\mathrm{e}.
\end{align}
The homogenized porous media region that models the trees is indicated as \(\Gamma_{\mathrm{tree}}\) (see Figure~\ref{fig:geometrical_setup}) and is used in one of the setups. 
The ground $\Gamma_{\mathrm{ground}}$ and $\Gamma_{\mathrm{wall}}$ are set to no-slip Dirichlet conditions, enforcing zero velocity at these surfaces, i.e.
\begin{align}
\bm{u}(\bm{x}, t) = \bm{0}, \quad \text{on } \Gamma_{k} \times (0,T], \quad \forall \Gamma_{k} \in \{\Gamma_\mathrm{ground}, \Gamma_\mathrm{wall}\} . 
\end{align}
We use a full-slip boundary condition for the fluid velocity at the open boundaries, i.e.
\begin{align}
\bm{u}(\bm{x}, t) \cdot \bm{n} = 0  & \quad \text{on } \Gamma_{k} \times (0,T], \quad \forall \Gamma_{k} \in \{\Gamma_\mathrm{sky}, \Gamma_{\mathrm{left}}, \Gamma_{\mathrm{right}}\}, 
\end{align}
where \(\bm{n}\) is the outward pointing normal vector. 
Specifically, a full slip boundary condition is applied at $\Gamma_{\mathrm{sky}}$, allowing tangential flow without friction. 
At the inflow $\Gamma_\mathrm{in}$ we apply a Dirichlet condition with an interpolated inlet velocity profile, while at the outflow $\Gamma_\mathrm{out}$ we use a fixed pressure boundary, allowing natural flow out of the domain. 
For the inlet $\Gamma_\mathrm{in}$, the same velocity profile as in~\cite{PASQUIER23} is used, depicted as
\begin{equation}
    u(z) = U_\mathrm{H}\left ( \frac{z}{z_\mathrm{H}} \right )^\alpha,
    \label{eq:u_profile}
\end{equation}
where $\alpha=0.3$, $U_{\mathrm{H}}$ is the inflow velocity at height $z_{\mathrm{H}}$, and $z_{\mathrm{H}}$ is the height of the walls. 
For the turbulence at the inlet, we used the vortex method according to Hettel~\textit{et al.}~\cite{HETTEL24}, where vortices are randomly predefined at the inlet. As in~\cite{GROMKE08,MERLIER18} Sulfur hexafluoride ($\text{SF}_6$) was used as a tracer gas and is emitted from $\Gamma_\mathrm{e}$ with a \(z\)-velocity of $U_{\mathrm{e}}$.
Thus, 
\begin{align}
\bm{u}(\bm{x}, t) = (0,0, U_{\mathrm{e}})^{\mathrm{T}}, \quad \text{on } \Gamma_{\mathrm{e}} \times (0,T]
\end{align}
and 
\begin{align}
C(\bm{x}, t) = C_{\text{SF}_{6}}, \quad \text{on } \Gamma_{\mathrm{e}} \times (0,T], 
\end{align}
where \(C_{\text{SF}_{6}}\) is the emitted concentration of \(\text{SF}_{6}\). 
The numerical parameters for the simulation, including the number of cells, are shown in Table~\ref{tab:numerical_parameters}.

\begin{table}[h!]
\centering
\caption{Physical and numerical parameters. Units are denoted in brackets.}
\resizebox{\textwidth}{!}{
\begin{tabular}{c c c c c c c c c}
\toprule
$Re \,[-]$ & $S\!c_{\mathrm{t}} \,[-]$ & $U_{\mathrm{H}} \,[\text{m}/\text{s}]$ & $\triangle x \,[\text{m}]$ & $\triangle t \,[\text{s}]$ & $U_e \,[\text{m}/\text{s}]$ &$T_{\mathrm{tot}} \,[\text{s}]$ & $H \,[\text{m}]$ & \#cells\,$[-]$ \\
\midrule
$37000$ & $0.3$ &$ 4.65$ &$2 \times 10^{-2}$ &$1.8\times 10^{-5}$ &$0.2054$&$25$&$0.12$ &$438.705\times10^6$\\
\bottomrule
\end{tabular}}
\label{tab:numerical_parameters}
\end{table}

In this setup, two lattices are used: one for the airflow distribution and one for the concentration distribution of the pollution. 
At specific boundary regions, we collide additionally with the chosen dynamics approximating the FBNSE or the ADE, respectively. This is particularly relevant at the inflow regions of the domain $\Gamma_\mathrm{in}$ and $\Gamma_\mathrm{e}$.
For the airflow distribution, we apply the collision according to  
\begin{equation}
    \forall \Gamma_{k} \in \{ \Gamma_{\mathrm{in}}, \Gamma_{\mathrm{e}}\} : \text{FBNSE \eqref{eq:fbnse}}.
\end{equation}
For the concentration distribution, we apply the collision according to 
\begin{align}
    \Gamma_{\mathrm{e}} : \text{ADE \eqref{eq:porousADE}}.
\end{align}
Further, for the boundary conditions for the concentration distribution are 
\begin{align}
    \bm{g} (\bm{x}, t) &= \bm{0}, \quad \text{on } \Gamma_{k} \times (0, T], \quad \forall \Gamma_{k} \in \{ \Gamma_{\mathrm{in}}, \Gamma_{\mathrm{out}}, \Gamma_{\mathrm{sky}}, \Gamma_{\mathrm{left}}, \Gamma_{\mathrm{right}}\}, \\
    C(\bm{x}, t) & = C_{0}, \quad \text{on } \Gamma_{k} \times (0, T], \quad \forall \Gamma_{k} \in \{\Gamma_{\mathrm{ground}}, \Gamma_{\mathrm{wall}}\},    
\end{align}
where \(C_{0}= \mathrm{const}\) is realized with standard bounce-back.

\subsubsection{Performance}
To assess the computational efficiency of the HLBM simulations, we measure performance in terms of mega lattice updates per second (MLUPs). The performance is evaluated under different configurations: with and without trees, and with and without a wall function (WF). The results provide insight into the computational cost associated with modeling urban flow scenarios and are shown in Table~\ref{tab:performance_canyon}.
\begin{table}[h!]
\centering
\caption{Computational performance in MLUPs for the configurations used in this work.}
\small
\begin{tabular}{l c c}
\toprule
Configuration & Without WF\,$[\text{MLUPs}]$ & With WF\,$[\text{MLUPs}]$ \\
\midrule
Tree-less canyon & $\sim 4600$ &$\sim 4100$ \\
Tree canyon & $\sim 4100$ & $\sim 2500$ \\
\bottomrule
\end{tabular}
\label{tab:performance_canyon}
\end{table}
It is observed that the porous ADE correction from \eqref{eq:correctionTerm} has a big impact on performance, which results in a loss of $\sim 500$ MLUPs, by adding the WF it is reduced even further. During this work, the WF is not optimized in its implementation and is therefore rather inefficient compared to the rest of the code.
The performance was investigated on a GPU node (4x NVIDIA H100 GPUs) on the Horeka cluster at KIT.

\subsubsection{Simulation results}\label{sec:sim_res}

A series of simulations were performed to evaluate airflow and pollutant dispersion in a street canyon, comparing the results against experimental data from Gromke~\textit{et al.}~\cite{GROMKE08}. Two configurations were studied: a street canyon without trees and one with a porous tree. Each scenario was simulated with and without a WF, following the methodology outlined by Guo~\textit{et al.}~\cite{GUO02,GUOZAHO02}. The Spalding wall function was employed to approximate equilibrium and non-equilibrium velocity profiles, with Dirichlet boundary conditions applied as described earlier. 
The concentration measured at the walls A and B was normalized according to
\begin{equation}
    c_\mathrm{+} = \frac{c_\mathrm{m} U_\mathrm{H} H}{Q_{\mathrm{SF}_6}/l},
\end{equation}
where $c_\mathrm{m}$ is the concentration, $Q_{\mathrm{SF}_6} = 1.359 \times10^{-6}\,\text{m}^3 / \text{s}$ is the volumetric flow rate of $SF_6$ and $l = 1.42\, \text{m}$ the length of the line sources.
The objective is to evaluate the usability of the above described numerical approach as a model for a DT of a city as well as to observe the impact of the WF.

\begin{figure}[ht]
    \subfloat[Experiment Gromke~\textit{et al.}~\cite{GROMKE08}]{
    \centering
    \begin{minipage}[c]{0.35\textwidth} %
        \centering
        \begin{minipage}[t]{1\textwidth}
        \begin{tikzpicture}
            \begin{axis}[
                axis on top,
                xlabel={$x/H$},
                ylabel={$z/H$},
                ylabel style={yshift=-10pt},
                width=1.3\textwidth, %
                xmin=-0.5, xmax=0.5,
                ymin=0, ymax=1.3,
                minor grid style={dotted},
                xtick={-0.4,-0.2,0,0.2,0.4},
                ytick={0,0.2,0.4,0.6,0.8,1,1.2},
                minor xtick={-0.3,-0.1,0.1,0.3},
                minor ytick={0.1,0.3,0.5,0.7,0.9,1.1},
                axis equal image,
            ]
                \addplot graphics[xmin=-0.5, xmax=0.5, ymin=0, ymax=1.3] {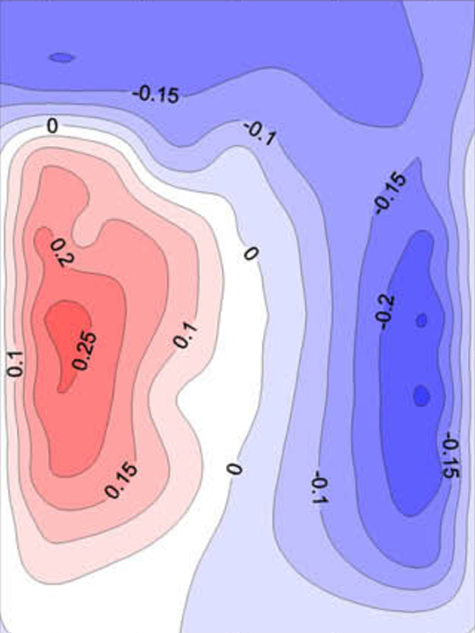};
            \end{axis}
        \end{tikzpicture}
        \end{minipage}
    \end{minipage}%
    }
    \begin{minipage}[c]{0.55\textwidth} %
        \centering
        \begin{minipage}[t]{0.45\textwidth}
        \subfloat[Merlier~\textit{et al.}~\cite{MERLIER18} without WF]{
            \centering
            \begin{tikzpicture}
                \begin{axis}[
                    axis on top,
                    xlabel={$x/H$},
                    ylabel={$z/H$},
                    ylabel style={yshift=-10pt},
                    width=1.8\textwidth, %
                    xmin=-0.5, xmax=0.5,
                    ymin=0, ymax=1.3,
                    minor grid style={dotted},
                    xtick={-0.4,-0.2,0,0.2,0.4},
                    ytick={0,0.2,0.4,0.6,0.8,1,1.2},
                    minor xtick={-0.3,-0.1,0.1,0.3},
                    minor ytick={0.1,0.3,0.5,0.7,0.9,1.1},
                    axis equal image,
                ]
                    \addplot graphics[xmin=-0.5, xmax=0.5, ymin=0, ymax=1.3] {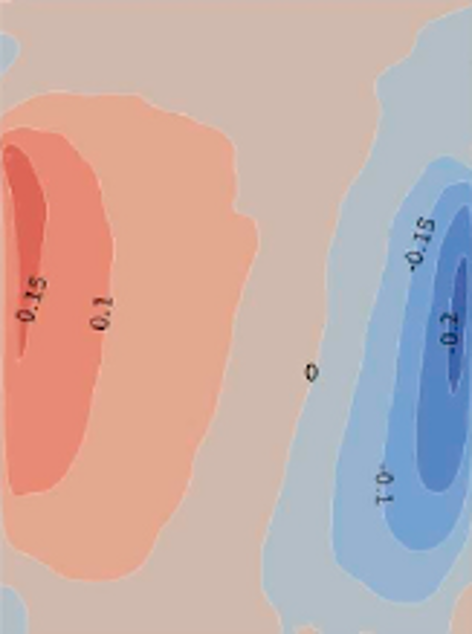};
                \end{axis}
            \end{tikzpicture}
        }
        \end{minipage}%
        \hfill
        \begin{minipage}[t]{0.45\textwidth}
            \subfloat[Ours without WF]{
            \centering
            \begin{tikzpicture}
                \begin{axis}[
                    axis on top,
                    xlabel={$x/H$},
                    ylabel={$z/H$},
                    ylabel style={yshift=-10pt},
                    width=1.8\textwidth, %
                    xmin=-0.5, xmax=0.5,
                    ymin=0, ymax=1.3,
                    minor grid style={dotted},
                    xtick={-0.4,-0.2,0,0.2,0.4},
                    ytick={0,0.2,0.4,0.6,0.8,1,1.2},
                    minor xtick={-0.3,-0.1,0.1,0.3},
                    minor ytick={0.1,0.3,0.5,0.7,0.9,1.1},
                    axis equal image,
                ]
                    \addplot graphics[xmin=-0.5, xmax=0.5, ymin=0, ymax=1.3] {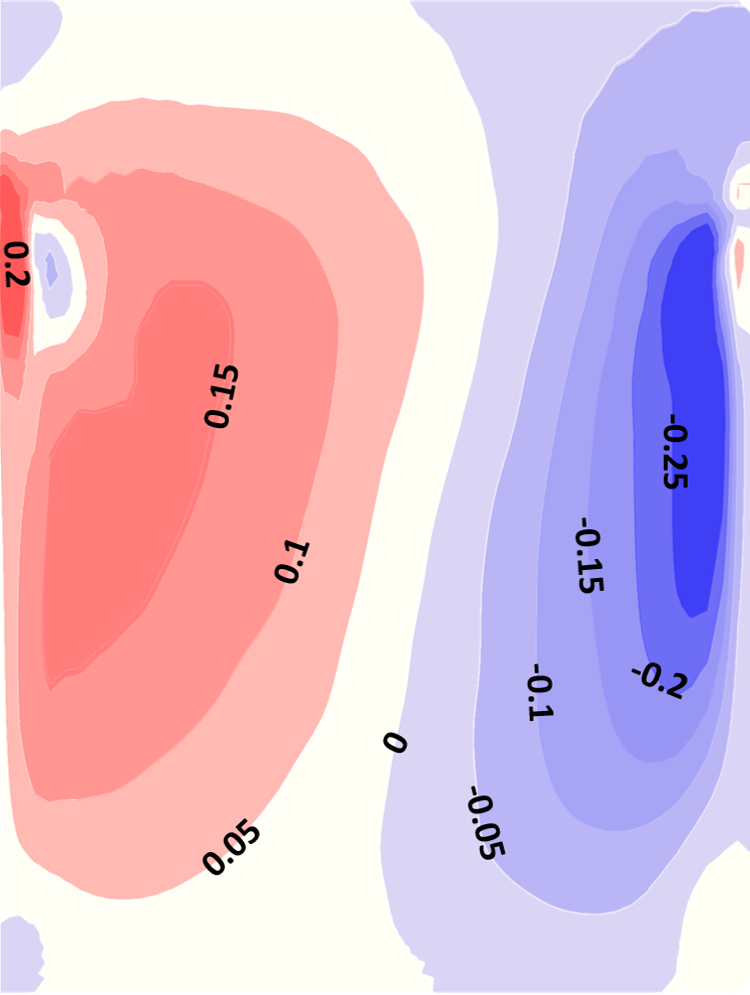};
                \end{axis}
            \end{tikzpicture}
            }
        \end{minipage}
        \\[1em]
        \begin{minipage}[t]{0.45\textwidth}
        \subfloat[Gromke~\textit{et al.}~\cite{GROMKE08} with WF]{
            \centering
            \begin{tikzpicture}
                \begin{axis}[
                    axis on top,
                    xlabel={$x/H$},
                    ylabel={$z/H$},
                    ylabel style={yshift=-10pt},
                    width=1.8\textwidth, %
                    xmin=-0.5, xmax=0.5,
                    ymin=0, ymax=1.3,
                    minor grid style={dotted},
                    xtick={-0.4,-0.2,0,0.2,0.4},
                    ytick={0,0.2,0.4,0.6,0.8,1,1.2},
                    minor xtick={-0.3,-0.1,0.1,0.3},
                    minor ytick={0.1,0.3,0.5,0.7,0.9,1.1},
                    axis equal image,
                ]
                    \addplot graphics[xmin=-0.5, xmax=0.5, ymin=0, ymax=1.3] {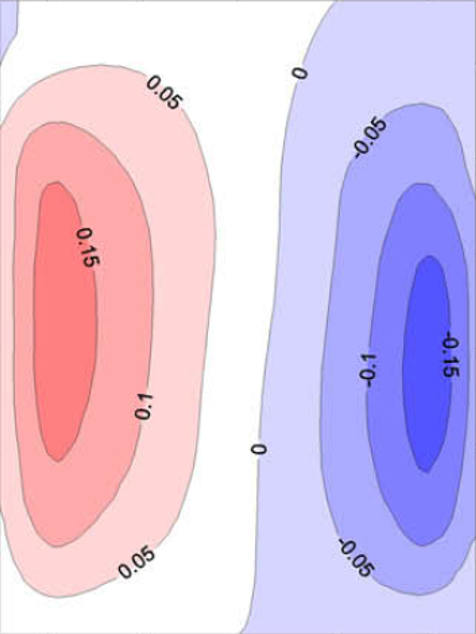};
                \end{axis}
            \end{tikzpicture}
            }
        \end{minipage}
        \hfill
        \begin{minipage}[t]{0.45\textwidth}
            \centering
        \subfloat[Ours with WF]{
            \begin{tikzpicture}
                \begin{axis}[
                    axis on top,
                    xlabel={$x/H$},
                    ylabel={$z/H$},
                    ylabel style={yshift=-10pt},
                    width=1.8\textwidth, %
                    xmin=-0.5, xmax=0.5,
                    ymin=0, ymax=1.3,
                    minor grid style={dotted},
                    xtick={-0.4,-0.2,0,0.2,0.4},
                    ytick={0,0.2,0.4,0.6,0.8,1,1.2},
                    minor xtick={-0.3,-0.1,0.1,0.3},
                    minor ytick={0.1,0.3,0.5,0.7,0.9,1.1},
                    axis equal image,
                ]
                    \addplot graphics[xmin=-0.5, xmax=0.5, ymin=0, ymax=1.3] {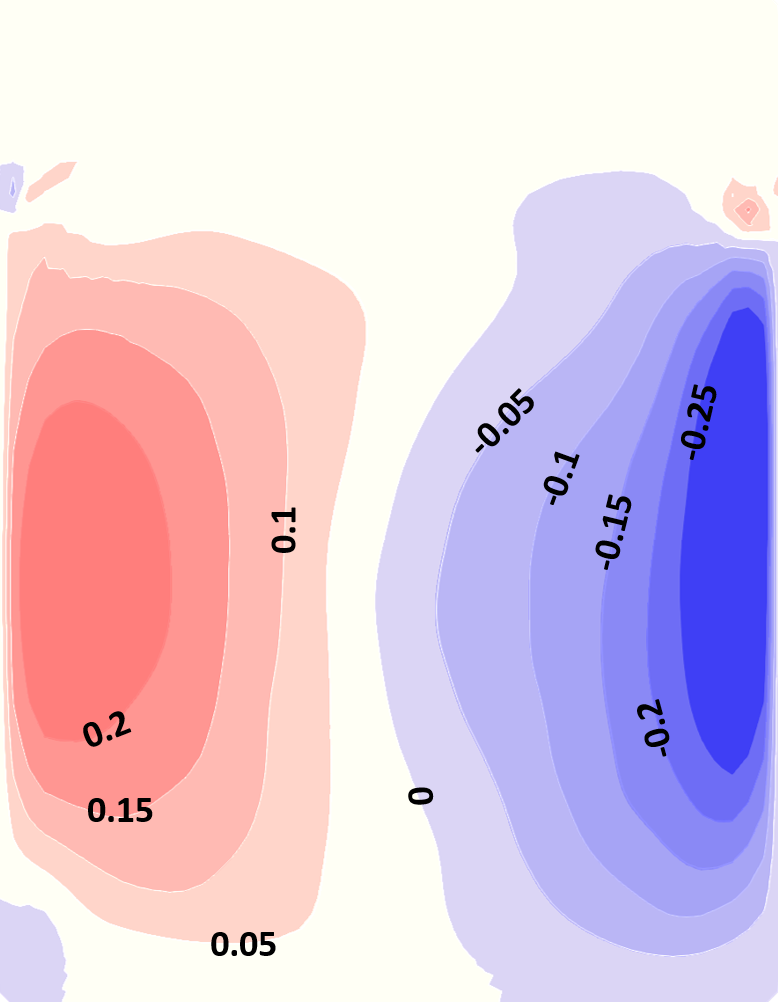};
                \end{axis}
            \end{tikzpicture}
        }
        \end{minipage}%
    \end{minipage}
    \caption{The experimental and simulation results for the normalized vertical velocity of a tree free canyon are shown. The experimental results from Gromke~\textit{et al.}~\cite{GROMKE08} are shown in (a), (b) shows numerical results from Merlier~\textit{et al.}~\cite{MERLIER18} without a WF, (c) shows our results without WF, (d) shows the numerical results of Gromke~\textit{et al.}~\cite{GROMKE08} with WF and (e) shows ours with a WF.}
    \label{fig:vel_comparison}
\end{figure}

\begin{figure}[htbp]
    \centering
    \subfloat[Experiment Gromke~\textit{et al.}~\cite{GROMKE08}]{
        \begin{tikzpicture}
            \begin{axis}[
                axis on top,
                xlabel={$x/H$},
                ylabel={$z/H$},
                xlabel style={font=\fontsize{7}{7}\selectfont,yshift=4pt},
                ylabel style={font=\fontsize{7}{7}\selectfont, yshift=-10pt},
                tick label style={font=\fontsize{7}{7}\selectfont},
                width=0.51\textwidth,
                xmin=-5, xmax=5,
                ymin=0, ymax=1,
                minor grid style={dotted},
                xtick={-5,-4,-3,-2,-1,0,1,2,3,4,5},
                ytick={0.5,1},
                axis equal image,
                tick align=outside,
            ]
                \addplot graphics[xmin=-5, xmax=5, ymin=0, ymax=1] {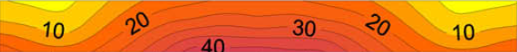};
            \end{axis}
        \end{tikzpicture}
    }
    
    \vspace{0.5em} %
    \subfloat[Merlier~\textit{et al.}~\cite{MERLIER18} without WF]{
        \begin{tikzpicture}
            \begin{axis}[
                axis on top,
                xlabel={$x/H$},
                ylabel={$z/H$},
                xlabel style={font=\fontsize{7}{7}\selectfont,yshift=4pt},
                ylabel style={font=\fontsize{7}{7}\selectfont, yshift=-10pt},
                tick label style={font=\fontsize{7}{7}\selectfont},
                width=0.51\textwidth,
                xmin=-5, xmax=5,
                ymin=0, ymax=1,
                minor grid style={dotted},
                xtick={-5,-4,-3,-2,-1,0,1,2,3,4,5},
                ytick={0.5,1},
                axis equal image,
                tick align=outside,
            ]
                \addplot graphics[xmin=-5, xmax=5, ymin=0, ymax=1] {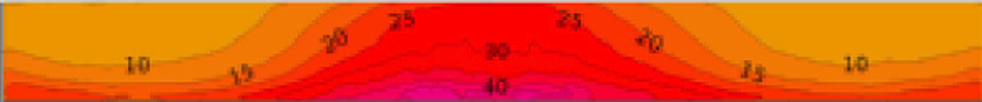};
            \end{axis}
        \end{tikzpicture}
    }
    \subfloat[Ours without WF]{
        \begin{tikzpicture}
            \begin{axis}[
                axis on top,
                xlabel={$x/H$},
                ylabel={$z/H$},
                xlabel style={font=\fontsize{7}{7}\selectfont,yshift=4pt},
                ylabel style={font=\fontsize{7}{7}\selectfont, yshift=-10pt},
                tick label style={font=\fontsize{7}{7}\selectfont},
                width=0.51\textwidth,
                xmin=-5, xmax=5,
                ymin=0, ymax=1,
                minor grid style={dotted},
                xtick={-5,-4,-3,-2,-1,0,1,2,3,4,5},
                ytick={0.5,1},
                axis equal image,
                tick align=outside,
            ]
                \addplot graphics[xmin=-5, xmax=5, ymin=0, ymax=1] {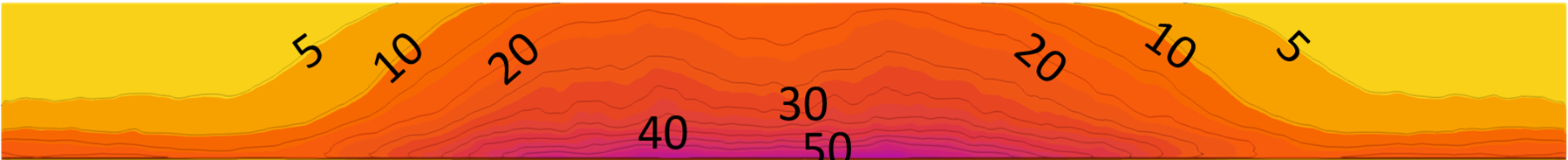};
            \end{axis}
        \end{tikzpicture}
    }

    \subfloat[Gromke~\textit{et al.}~\cite{GROMKE08} with WF]{
        \begin{tikzpicture}
            \begin{axis}[
                axis on top,
                xlabel={$x/H$},
                ylabel={$z/H$},
                xlabel style={font=\fontsize{7}{7}\selectfont,yshift=4pt},
                ylabel style={font=\fontsize{7}{7}\selectfont, yshift=-10pt},
                tick label style={font=\fontsize{7}{7}\selectfont},
                width=0.51\textwidth,
                xmin=-5, xmax=5,
                ymin=0, ymax=1,
                minor grid style={dotted},
                xtick={-5,-4,-3,-2,-1,0,1,2,3,4,5},
                ytick={0.5,1},
                axis equal image,
                tick align=outside,
            ]
                \addplot graphics[xmin=-5, xmax=5, ymin=0, ymax=1] {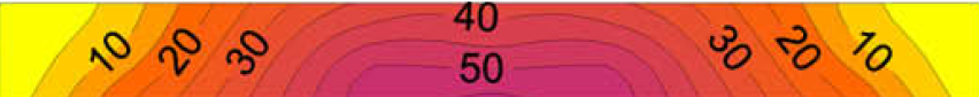};
            \end{axis}
        \end{tikzpicture}
    }
    \subfloat[Ours with WF]{
        \begin{tikzpicture}
            \begin{axis}[
                axis on top,
                xlabel={$x/H$},
                ylabel={$z/H$},
                xlabel style={font=\fontsize{7}{7}\selectfont,yshift=4pt},
                ylabel style={font=\fontsize{7}{7}\selectfont, yshift=-10pt},
                tick label style={font=\fontsize{7}{7}\selectfont},
                width=0.51\textwidth,
                xmin=-5, xmax=5,
                ymin=0, ymax=1,
                minor grid style={dotted},
                xtick={-5,-4,-3,-2,-1,0,1,2,3,4,5},
                ytick={0.5,1},
                axis equal image,
                tick align=outside,
            ]
                \addplot graphics[xmin=-5, xmax=5, ymin=0, ymax=1] {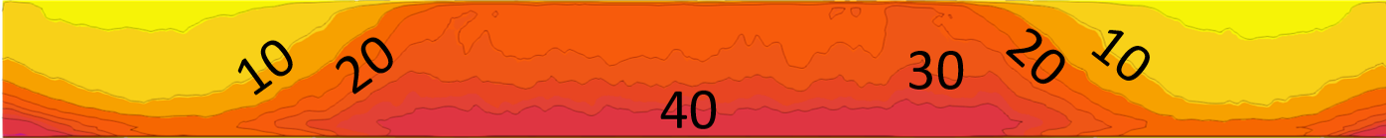};
            \end{axis}
        \end{tikzpicture}
    }
    \caption{Normalized concentration $c_+$ at the Wall A. (a) shows the experimental results according to Gromke~\textit{et al.}~\cite{GROMKE08}, (b)  the results without a WF from Merlier~\textit{et al.}~\cite{MERLIER18}, (c) ours without a WF, (d) results from Gromke~\textit{et al.}~\cite{GROMKE08} with a WF and (e) shows our with WF}
    \label{fig:concentration_wallA_no_tree}
\end{figure}

\begin{figure}[htbp]
    \centering
    \subfloat[Experiment Gromke~\textit{et al.}~\cite{GROMKE08}]{
        \begin{tikzpicture}
            \begin{axis}[
                axis on top,
                xlabel={$x/H$},
                ylabel={$z/H$},
                xlabel style={font=\fontsize{7}{7}\selectfont,yshift=4pt},
                ylabel style={font=\fontsize{7}{7}\selectfont, yshift=-10pt},
                tick label style={font=\fontsize{7}{7}\selectfont},
                width=0.51\textwidth, %
                xmin=-5, xmax=5,
                ymin=0, ymax=1,
                minor grid style={dotted},
                xtick={-5,-4,-3,-2,-1,0,1,2,3,4,5},
                ytick={0.5,1},
                axis equal image,
                tick align=outside,
            ]
                \addplot graphics[xmin=-5, xmax=5, ymin=0, ymax=1] {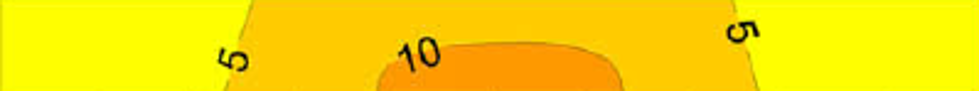};
            \end{axis}
        \end{tikzpicture}
    }
    
    \vspace{0.5em} %
    \subfloat[Merlier~\textit{et al.}~\cite{MERLIER18} without WF]{
        \begin{tikzpicture}
            \begin{axis}[
                axis on top,
                xlabel={$x/H$},
                ylabel={$z/H$},
                xlabel style={font=\fontsize{7}{7}\selectfont,yshift=4pt},
                ylabel style={font=\fontsize{7}{7}\selectfont, yshift=-10pt},
                tick label style={font=\fontsize{7}{7}\selectfont},
                width=0.51\textwidth,
                xmin=-5, xmax=5,
                ymin=0, ymax=1,
                minor grid style={dotted},
                xtick={-5,-4,-3,-2,-1,0,1,2,3,4,5},
                ytick={0.5,1},
                axis equal image,
                tick align=outside,
            ]
                \addplot graphics[xmin=-5, xmax=5, ymin=0, ymax=1] {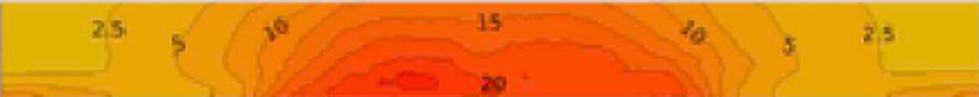};
            \end{axis}
        \end{tikzpicture}
    }
    \subfloat[Ours without WF]{
        \begin{tikzpicture}
            \begin{axis}[
                axis on top,
                xlabel={$x/H$},
                ylabel={$z/H$},
                xlabel style={font=\fontsize{7}{7}\selectfont,yshift=4pt},
                ylabel style={font=\fontsize{7}{7}\selectfont, yshift=-10pt},
                tick label style={font=\fontsize{7}{7}\selectfont},
                width=0.51\textwidth,
                xmin=-5, xmax=5,
                ymin=0, ymax=1,
                minor grid style={dotted},
                xtick={-5,-4,-3,-2,-1,0,1,2,3,4,5},
                ytick={0.5,1},
                axis equal image,
                tick align=outside,
            ]
                \addplot graphics[xmin=-5, xmax=5, ymin=0, ymax=1] {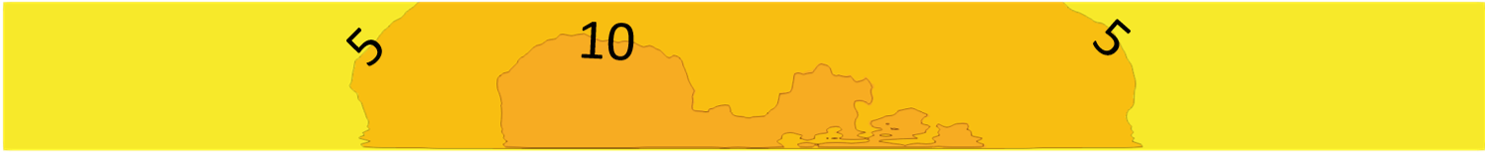};
            \end{axis}
        \end{tikzpicture}
    }

    \subfloat[Gromke~\textit{et al.}~\cite{GROMKE08} with WF]{
        \begin{tikzpicture}
            \begin{axis}[
                axis on top,
                xlabel={$x/H$},
                ylabel={$z/H$},
                xlabel style={font=\fontsize{7}{7}\selectfont,yshift=4pt},
                ylabel style={font=\fontsize{7}{7}\selectfont, yshift=-10pt},
                tick label style={font=\fontsize{7}{7}\selectfont},
                width=0.51\textwidth,
                xmin=-5, xmax=5,
                ymin=0, ymax=1,
                minor grid style={dotted},
                xtick={-5,-4,-3,-2,-1,0,1,2,3,4,5},
                ytick={0.5,1},
                axis equal image,
                tick align=outside,
            ]
                \addplot graphics[xmin=-5, xmax=5, ymin=0, ymax=1] {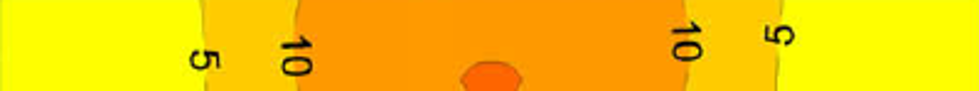};
            \end{axis}
        \end{tikzpicture}
    }
    \subfloat[Ours with WF]{
        \begin{tikzpicture}
            \begin{axis}[
                axis on top,
                xlabel={$x/H$},
                ylabel={$z/H$},
                xlabel style={font=\fontsize{7}{7}\selectfont,yshift=4pt},
                ylabel style={font=\fontsize{7}{7}\selectfont, yshift=-10pt},
                tick label style={font=\fontsize{7}{7}\selectfont},
                width=0.51\textwidth,
                xmin=-5, xmax=5,
                ymin=0, ymax=1,
                minor grid style={dotted},
                xtick={-5,-4,-3,-2,-1,0,1,2,3,4,5},
                ytick={0.5,1},
                axis equal image,
                tick align=outside,
            ]
                \addplot graphics[xmin=-5, xmax=5, ymin=0, ymax=1] {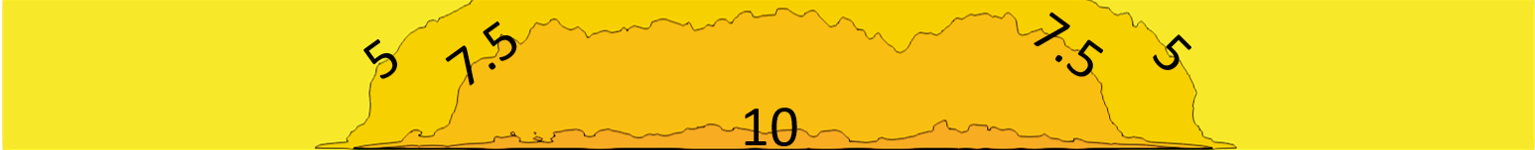};
            \end{axis}
        \end{tikzpicture}
    }
    \caption{Normalized concentration $c_+$ at the Wall B. (a) shows the experimental results according to Gromke~\textit{et al.}~\cite{GROMKE08}, (b)  the results without a WF from Merlier~\textit{et al.}~\cite{MERLIER18}, (c) ours without a WF, (d) results from Gromke~\textit{et al.}~\cite{GROMKE08} with a WF and (e) shows our with WF}
    \label{fig:concentration_wallB_no_tree}
\end{figure}

\textbf{Tree-free canyon:} Figure~\ref{fig:vel_comparison} presents the normalized vertical velocity profiles for a tree-free canyon. The subfigures compare the experimental data with numerical results from other studies and the current study, both with and without a wall function. In Figure~\ref{fig:vel_comparison}~(a), the experimental results of Gromke~\textit{et al.}~\cite{GROMKE08} provide the baseline for comparison. Subfigures (b) and (c) illustrate the results of simulations without a WF. Although the general velocity profiles are consistent with the experimental data, discrepancies are observed at the edges of the wall, where the maximum vertical velocities are overestimated. 
Figures~\ref{fig:vel_comparison}~(d) and (e) incorporate a WF in the simulations. Here, the maximum vertical velocity is shifted toward the center of the canyon, aligning more closely with the experimental observations. Figure~\ref{fig:vel_comparison}~(e) demonstrates particularly strong agreement with the experimental data, validating the effectiveness of the wall model in capturing the critical flow characteristics in urban environments.
The pollutant concentration distribution is further analyzed in Figures~\ref{fig:concentration_wallA_no_tree} and \ref{fig:concentration_wallB_no_tree}, which show normalized concentrations along walls A and B, respectively, in the absence of trees. In Figure~\ref{fig:concentration_wallA_no_tree}~(a), experimental data from Gromke~\textit{et al.}~\cite{GROMKE08} illustrate the expected pollutant accumulation patterns. Subfigures (b) and (c) present numerical results without a WF. Although the general trends and stratification of the concentration layers align with the experimental data, the pollutant accumulation zones are narrower and exhibit higher gradients. This behavior is attributed to the elevated vertical velocities near the edges of the wall, as seen in Figures~\ref{fig:vel_comparison}~(b) and (c). The thinner pollutant layers near the walls indicate an over-prediction of upward pollutant transport.
When using a WF, as shown in Figures~\ref{fig:concentration_wallA_no_tree}~(d) and (e), the pollutant accumulation patterns show improved agreement with the experimental results. The concentration layers are wider and better reflect the observed stratification, particularly along the central portions of the wall. This improvement corresponds to the more realistic velocity profiles shown in Figures~\ref{fig:vel_comparison}~(d) and (e), where the maximum vertical velocities are centered and reduced near the walls.


Along wall B depicted in Figure~\ref{fig:concentration_wallB_no_tree} the results without a WF differ strongly. The reference case in (b) shows stratified layers with concentrations higher than those of the experiment data in (a), while our results in (c) show strong alignment with (a). The results with a WF in (d) and (e) also agree with (a).

\begin{figure}[ht]
    \subfloat[Experiment Gromke~\textit{et al.}~\cite{GROMKE08}]{
    \centering
    \begin{minipage}[c]{0.35\textwidth} %
        \centering
        \begin{minipage}[t]{1\textwidth}
        \begin{tikzpicture}
            \begin{axis}[
                axis on top,
                xlabel={$x/H$},
                ylabel={$z/H$},
                ylabel style={yshift=-10pt},
                width=1.3\textwidth, %
                xmin=-0.5, xmax=0.5,
                ymin=0, ymax=1.3,
                minor grid style={dotted},
                xtick={-0.4,-0.2,0,0.2,0.4},
                ytick={0,0.2,0.4,0.6,0.8,1,1.2},
                minor xtick={-0.3,-0.1,0.1,0.3},
                minor ytick={0.1,0.3,0.5,0.7,0.9,1.1},
                axis equal image,
            ]
                \addplot graphics[xmin=-0.5, xmax=0.5, ymin=0, ymax=1.3] {vel_tree_m.png};
            \end{axis}
        \end{tikzpicture}
        \end{minipage}
    \end{minipage}%
    }
    \begin{minipage}[c]{0.55\textwidth} %
        \centering
        \begin{minipage}[t]{0.45\textwidth}
        \subfloat[Merlier~\textit{et al.}~\cite{MERLIER18} without WF]{
            \centering
            \begin{tikzpicture}
                \begin{axis}[
                    axis on top,
                    xlabel={$x/H$},
                    ylabel={$z/H$},
                    ylabel style={yshift=-10pt},
                    width=1.8\textwidth, %
                    xmin=-0.5, xmax=0.5,
                    ymin=0, ymax=1.3,
                    minor grid style={dotted},
                    xtick={-0.4,-0.2,0,0.2,0.4},
                    ytick={0,0.2,0.4,0.6,0.8,1,1.2},
                    minor xtick={-0.3,-0.1,0.1,0.3},
                    minor ytick={0.1,0.3,0.5,0.7,0.9,1.1},
                    axis equal image,
                ]
                    \addplot graphics[xmin=-0.5, xmax=0.5, ymin=0, ymax=1.3] {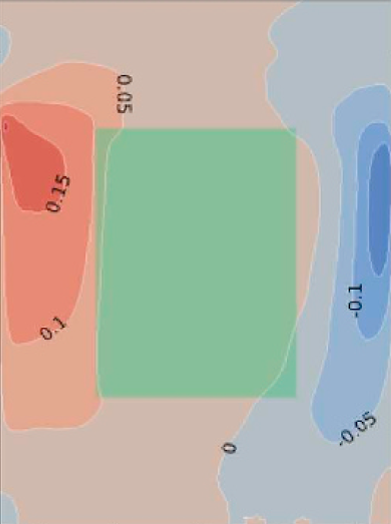};
                \end{axis}
            \end{tikzpicture}
        }
        \end{minipage}%
        \hfill
        \begin{minipage}[t]{0.45\textwidth}
            \subfloat[Ours without WF]{
            \centering
            \begin{tikzpicture}
                \begin{axis}[
                    axis on top,
                    xlabel={$x/H$},
                    ylabel={$z/H$},
                    ylabel style={yshift=-10pt},
                    width=1.8\textwidth, %
                    xmin=-0.5, xmax=0.5,
                    ymin=0, ymax=1.3,
                    minor grid style={dotted},
                    xtick={-0.4,-0.2,0,0.2,0.4},
                    ytick={0,0.2,0.4,0.6,0.8,1,1.2},
                    minor xtick={-0.3,-0.1,0.1,0.3},
                    minor ytick={0.1,0.3,0.5,0.7,0.9,1.1},
                    axis equal image,
                ]
                    \addplot graphics[xmin=-0.5, xmax=0.5, ymin=0, ymax=1.3] {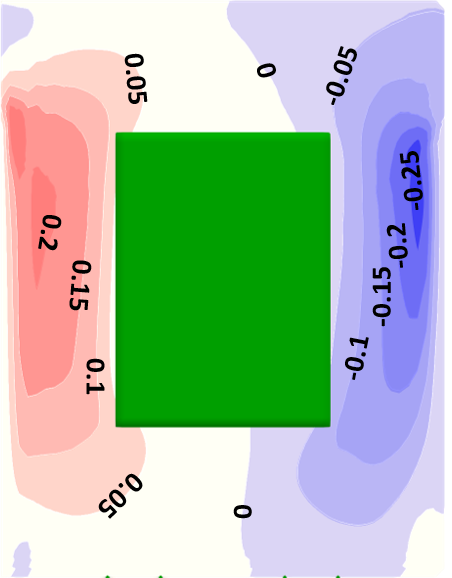};
                \end{axis}
            \end{tikzpicture}
            }
        \end{minipage}
        \\[1em]
        \begin{minipage}[t]{0.45\textwidth}
        \subfloat[Gromke~\textit{et al.}~\cite{GROMKE08} with WF]{
            \centering
            \begin{tikzpicture}
                \begin{axis}[
                    axis on top,
                    xlabel={$x/H$},
                    ylabel={$z/H$},
                    ylabel style={yshift=-10pt},
                    width=1.8\textwidth, %
                    xmin=-0.5, xmax=0.5,
                    ymin=0, ymax=1.3,
                    minor grid style={dotted},
                    xtick={-0.4,-0.2,0,0.2,0.4},
                    ytick={0,0.2,0.4,0.6,0.8,1,1.2},
                    minor xtick={-0.3,-0.1,0.1,0.3},
                    minor ytick={0.1,0.3,0.5,0.7,0.9,1.1},
                    axis equal image,
                ]
                    \addplot graphics[xmin=-0.5, xmax=0.5, ymin=0, ymax=1.3] {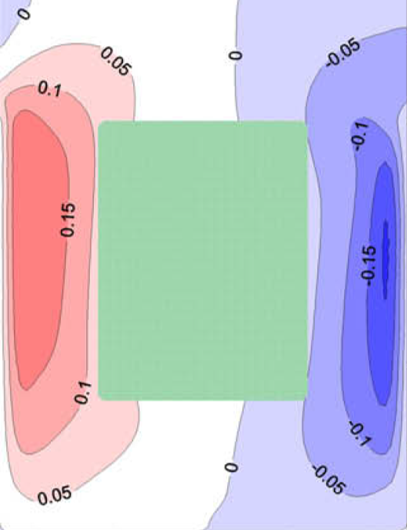};
                \end{axis}
            \end{tikzpicture}
            }
        \end{minipage}
        \hfill
        \begin{minipage}[t]{0.45\textwidth}
            \centering
        \subfloat[Ours with WF]{
            \begin{tikzpicture}
                \begin{axis}[
                    axis on top,
                    xlabel={$x/H$},
                    ylabel={$z/H$},
                    ylabel style={yshift=-10pt},
                    width=1.8\textwidth, %
                    xmin=-0.5, xmax=0.5,
                    ymin=0, ymax=1.3,
                    minor grid style={dotted},
                    xtick={-0.4,-0.2,0,0.2,0.4},
                    ytick={0,0.2,0.4,0.6,0.8,1,1.2},
                    minor xtick={-0.3,-0.1,0.1,0.3},
                    minor ytick={0.1,0.3,0.5,0.7,0.9,1.1},
                    axis equal image,
                ]
                    \addplot graphics[xmin=-0.5, xmax=0.5, ymin=0, ymax=1.3] {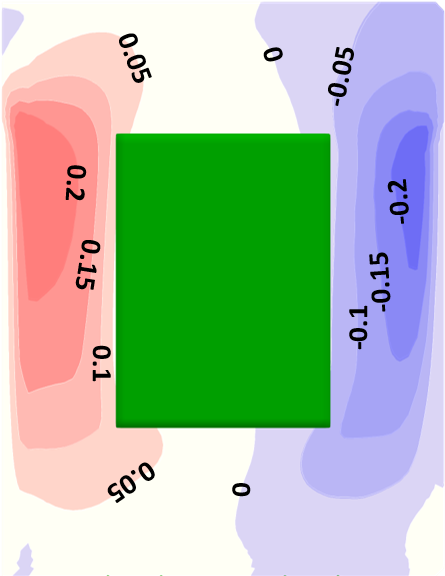};
                \end{axis}
            \end{tikzpicture}
        }
        \end{minipage}%
    \end{minipage}
    \caption{The experimental and simulation results for the normalized vertical velocity of a  canyon with a tree are shown. The experimental results from Gromke~\textit{et al.}~\cite{GROMKE08} are shown in (a), (b) shows numerical results from Merlier~\textit{et al.}~\cite{MERLIER18} without a WF, (c) shows our results without WF, (d) shows the numerical results of Gromke~\textit{et al.}~\cite{GROMKE08} with WF and (e) shows ours with a WF.}
    \label{fig:vel_comparison_tree}
\end{figure}
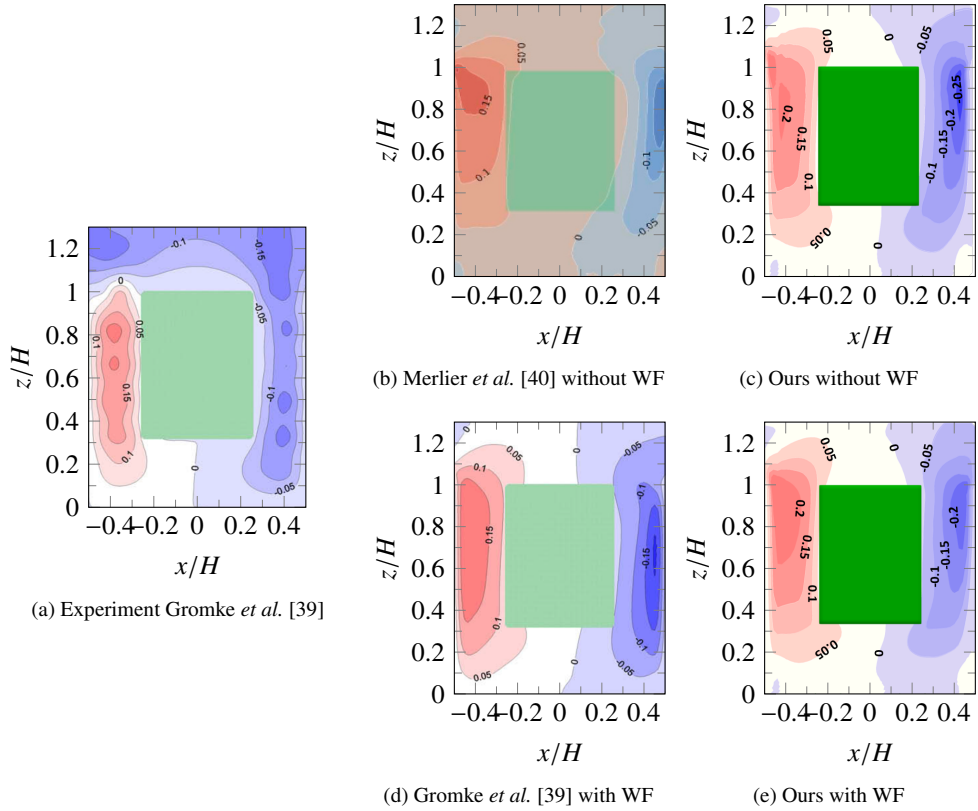

\begin{figure}[htbp]
    \centering
    \subfloat[Experiment Gromke~\textit{et al.}~\cite{GROMKE08}]{
        \begin{tikzpicture}
            \begin{axis}[
                axis on top,
                xlabel={$x/H$},
                ylabel={$z/H$},
                xlabel style={font=\fontsize{7}{7}\selectfont,yshift=4pt},
                ylabel style={font=\fontsize{7}{7}\selectfont, yshift=-10pt},
                tick label style={font=\fontsize{7}{7}\selectfont},
                width=0.51\textwidth, %
                xmin=-5, xmax=5,
                ymin=0, ymax=1,
                minor grid style={dotted},
                xtick={-5,-4,-3,-2,-1,0,1,2,3,4,5},
                ytick={0.5,1},
                axis equal image,
                tick align=outside,
            ]
                \addplot graphics[xmin=-5, xmax=5, ymin=0, ymax=1] {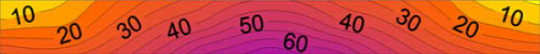};
            \end{axis}
        \end{tikzpicture}
    }
    
    \vspace{0.5em} %
    \subfloat[Merlier~\textit{et al.}~\cite{MERLIER18} without WF]{
        \begin{tikzpicture}
            \begin{axis}[
                axis on top,
                xlabel={$x/H$},
                ylabel={$z/H$},
                xlabel style={font=\fontsize{7}{7}\selectfont,yshift=4pt},
                ylabel style={font=\fontsize{7}{7}\selectfont, yshift=-10pt},
                tick label style={font=\fontsize{7}{7}\selectfont},
                width=0.51\textwidth,
                xmin=-5, xmax=5,
                ymin=0, ymax=1,
                minor grid style={dotted},
                xtick={-5,-4,-3,-2,-1,0,1,2,3,4,5},
                ytick={0.5,1},
                axis equal image,
                tick align=outside,
            ]
                \addplot graphics[xmin=-5, xmax=5, ymin=0, ymax=1] {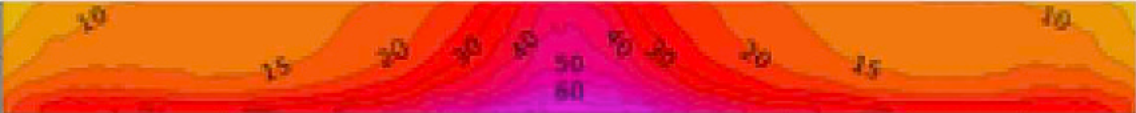};
            \end{axis}
        \end{tikzpicture}
    }
    \subfloat[Ours without WF]{
        \begin{tikzpicture}
            \begin{axis}[
                axis on top,
                xlabel={$x/H$},
                ylabel={$z/H$},
                xlabel style={font=\fontsize{7}{7}\selectfont,yshift=4pt},
                ylabel style={font=\fontsize{7}{7}\selectfont, yshift=-10pt},
                tick label style={font=\fontsize{7}{7}\selectfont},
                width=0.51\textwidth,
                xmin=-5, xmax=5,
                ymin=0, ymax=1,
                minor grid style={dotted},
                xtick={-5,-4,-3,-2,-1,0,1,2,3,4,5},
                ytick={0.5,1},
                axis equal image,
                tick align=outside,
            ]
                \addplot graphics[xmin=-5, xmax=5, ymin=0, ymax=1] {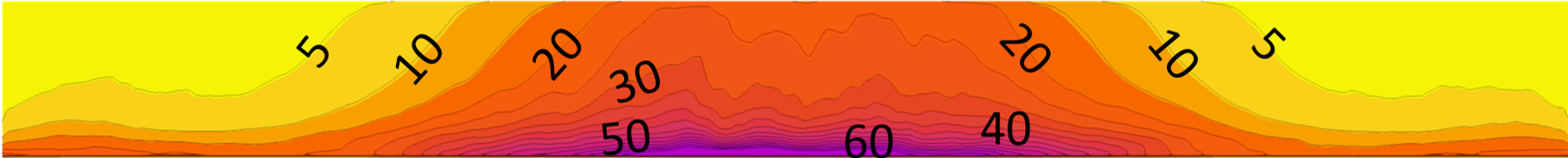};
            \end{axis}
        \end{tikzpicture}
    }

    \subfloat[Gromke~\textit{et al.}~\cite{GROMKE08} with WF]{
        \begin{tikzpicture}
            \begin{axis}[
                axis on top,
                xlabel={$x/H$},
                ylabel={$z/H$},
                xlabel style={font=\fontsize{7}{7}\selectfont,yshift=4pt},
                ylabel style={font=\fontsize{7}{7}\selectfont, yshift=-10pt},
                tick label style={font=\fontsize{7}{7}\selectfont},
                width=0.51\textwidth,
                xmin=-5, xmax=5,
                ymin=0, ymax=1,
                minor grid style={dotted},
                xtick={-5,-4,-3,-2,-1,0,1,2,3,4,5},
                ytick={0.5,1},
                axis equal image,
                tick align=outside,
            ]
                \addplot graphics[xmin=-5, xmax=5, ymin=0, ymax=1] {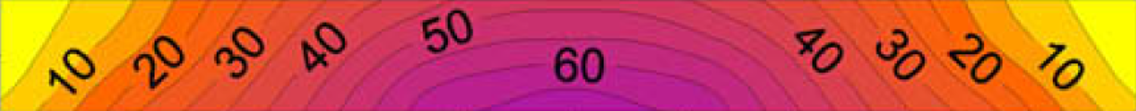};
            \end{axis}
        \end{tikzpicture}
    }
    \subfloat[Ours with WF]{
        \begin{tikzpicture}
            \begin{axis}[
                axis on top,
                xlabel={$x/H$},
                ylabel={$z/H$},
                xlabel style={font=\fontsize{7}{7}\selectfont,yshift=4pt},
                ylabel style={font=\fontsize{7}{7}\selectfont, yshift=-10pt},
                tick label style={font=\fontsize{7}{7}\selectfont},
                width=0.51\textwidth,
                xmin=-5, xmax=5,
                ymin=0, ymax=1,
                minor grid style={dotted},
                xtick={-5,-4,-3,-2,-1,0,1,2,3,4,5},
                ytick={0.5,1},
                axis equal image,
                tick align=outside,
            ]
                \addplot graphics[xmin=-5, xmax=5, ymin=0, ymax=1] {conz_wallA_tree_wf.png};
            \end{axis}
        \end{tikzpicture}
    }
    \caption{Normalized concentration $c_+$ at the Wall A for a canyon with a tree. (a) shows the experimental results according to Gromke~\textit{et al.}~\cite{GROMKE08}, (b)  the results without a WF from Merlier~\textit{et al.}~\cite{MERLIER18}, (c) ours without a WF, (d) results from Gromke~\textit{et al.}~\cite{GROMKE08} with a WF and (e) shows our with WF}
    \label{fig:concentration_wallA_tree}
\end{figure}

\begin{figure}[htbp]
    \centering
    \subfloat[Experiment Gromke~\textit{et al.}~\cite{GROMKE08}]{
        \begin{tikzpicture}
            \begin{axis}[
                axis on top,
                xlabel={$x/H$},
                ylabel={$z/H$},
                xlabel style={font=\fontsize{7}{7}\selectfont,yshift=4pt},
                ylabel style={font=\fontsize{7}{7}\selectfont, yshift=-10pt},
                tick label style={font=\fontsize{7}{7}\selectfont},
                width=0.51\textwidth, %
                xmin=-5, xmax=5,
                ymin=0, ymax=1,
                minor grid style={dotted},
                xtick={-5,-4,-3,-2,-1,0,1,2,3,4,5},
                ytick={0.5,1},
                axis equal image,
                tick align=outside,
            ]
                \addplot graphics[xmin=-5, xmax=5, ymin=0, ymax=1] {conz_wallB_tree_exp.png};
            \end{axis}
        \end{tikzpicture}
    }
    
    \vspace{0.5em} %
    \subfloat[Merlier~\textit{et al.}~\cite{MERLIER18} without WF]{
        \begin{tikzpicture}
            \begin{axis}[
                axis on top,
                xlabel={$x/H$},
                ylabel={$z/H$},
                xlabel style={font=\fontsize{7}{7}\selectfont,yshift=4pt},
                ylabel style={font=\fontsize{7}{7}\selectfont, yshift=-10pt},
                tick label style={font=\fontsize{7}{7}\selectfont},
                width=0.51\textwidth,
                xmin=-5, xmax=5,
                ymin=0, ymax=1,
                minor grid style={dotted},
                xtick={-5,-4,-3,-2,-1,0,1,2,3,4,5},
                ytick={0.5,1},
                axis equal image,
                tick align=outside,
            ]
                \addplot graphics[xmin=-5, xmax=5, ymin=0, ymax=1] {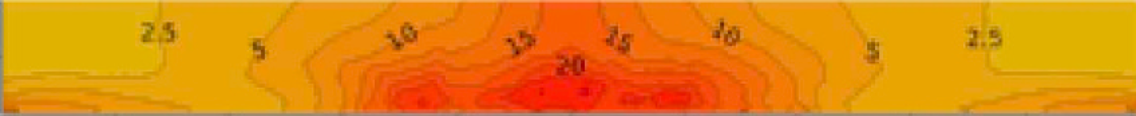};
            \end{axis}
        \end{tikzpicture}
    }
    \subfloat[Ours without WF]{
        \begin{tikzpicture}
            \begin{axis}[
                axis on top,
                xlabel={$x/H$},
                ylabel={$z/H$},
                xlabel style={font=\fontsize{7}{7}\selectfont,yshift=4pt},
                ylabel style={font=\fontsize{7}{7}\selectfont, yshift=-10pt},
                tick label style={font=\fontsize{7}{7}\selectfont},
                width=0.51\textwidth,
                xmin=-5, xmax=5,
                ymin=0, ymax=1,
                minor grid style={dotted},
                xtick={-5,-4,-3,-2,-1,0,1,2,3,4,5},
                ytick={0.5,1},
                axis equal image,
                tick align=outside,
            ]
                \addplot graphics[xmin=-5, xmax=5, ymin=0, ymax=1] {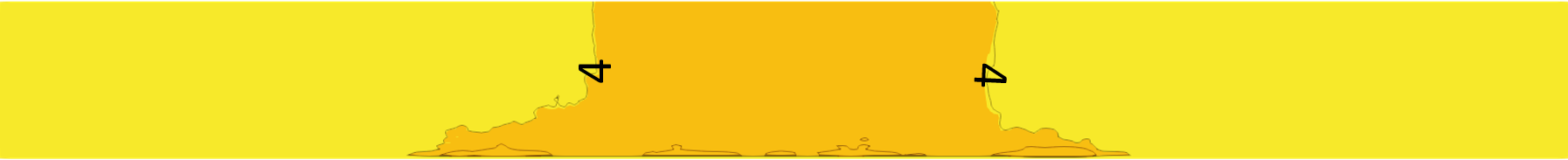};
            \end{axis}
        \end{tikzpicture}
    }

    \subfloat[Gromke~\textit{et al.}~\cite{GROMKE08} with WF]{
        \begin{tikzpicture}
            \begin{axis}[
                axis on top,
                xlabel={$x/H$},
                ylabel={$z/H$},
                xlabel style={font=\fontsize{7}{7}\selectfont,yshift=4pt},
                ylabel style={font=\fontsize{7}{7}\selectfont, yshift=-10pt},
                tick label style={font=\fontsize{7}{7}\selectfont},
                width=0.51\textwidth,
                xmin=-5, xmax=5,
                ymin=0, ymax=1,
                minor grid style={dotted},
                xtick={-5,-4,-3,-2,-1,0,1,2,3,4,5},
                ytick={0.5,1},
                axis equal image,
                tick align=outside,
            ]
                \addplot graphics[xmin=-5, xmax=5, ymin=0, ymax=1] {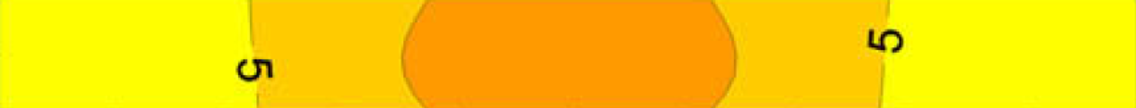};
            \end{axis}
        \end{tikzpicture}
    }
    \subfloat[Ours with WF]{
        \begin{tikzpicture}
            \begin{axis}[
                axis on top,
                xlabel={$x/H$},
                ylabel={$z/H$},
                xlabel style={font=\fontsize{7}{7}\selectfont,yshift=4pt},
                ylabel style={font=\fontsize{7}{7}\selectfont, yshift=-10pt},
                tick label style={font=\fontsize{7}{7}\selectfont},
                width=0.51\textwidth,
                xmin=-5, xmax=5,
                ymin=0, ymax=1,
                minor grid style={dotted},
                xtick={-5,-4,-3,-2,-1,0,1,2,3,4,5},
                ytick={0.5,1},
                axis equal image,
                tick align=outside,
            ]
                \addplot graphics[xmin=-5, xmax=5, ymin=0, ymax=1] {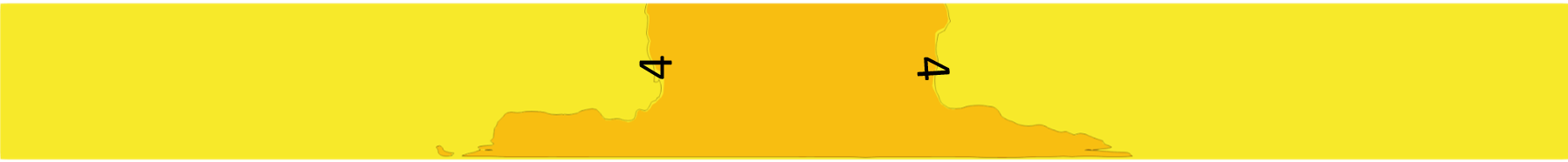};
            \end{axis}
        \end{tikzpicture}
    }
    \caption{Normalized concentration $c_+$ at the Wall B for a canyon with a tree. (a) shows the experimental results according to Gromke~\textit{et al.}~\cite{GROMKE08}, (b)  the results without a WF from Merlier~\textit{et al.}~\cite{MERLIER18}, (c) ours without a WF, (d) results from Gromke~\textit{et al.}~\cite{GROMKE08} with a WF and (e) shows our with WF}
    \label{fig:concentration_wallB_tree}
\end{figure}

\textbf{Tree-canyon: } Similar to the tree-free canyon case, Figure~\ref{fig:vel_comparison_tree} presents the simulation results for the velocity distribution in a street canyon with a porous tree. These results are compared against experimental data from Gromke~\textit{et al.}~\cite{GROMKE08} (Figure~\ref{fig:vel_comparison_tree}~(a)) and numerical results from Merlier~\textit{et al.}~\cite{MERLIER18} (Figure~\ref{fig:vel_comparison_tree}~(b)), as well as reference simulations with and without WF.
Figures~\ref{fig:vel_comparison_tree}~(c) and (e), which represent our numerical results without and with WF, respectively, show a similar overall velocity distribution to the experimental data in (a). However, slight differences can be observed, particularly in the velocity magnitude near the walls. Without WF (c), the velocity distribution captures the general pattern observed in (a) but exhibits some deviations in the center and near the edges of the canyon. When WF is included (e), the results remain largely consistent with (c), suggesting that WF has a minimal impact in this setup due to the narrow spacing between the walls and the porous tree zone.
Comparison with Merlier~\textit{et al.}~\cite{MERLIER18} (b) and the numerical results of Gromke~\textit{et al.}~\cite{GROMKE08} (d) shows that our simulations align well with previous numerical approaches, reinforcing the validity of our model.
Figure~\ref{fig:concentration_wallA_tree} presents the normalized concentration $c_\mathrm{+}$ along wall A, while Figure~\ref{fig:concentration_wallB_tree} shows the concentration along wall B for a canyon with a tree. The results are compared against experimental data from Gromke~\textit{et al.}~\cite{GROMKE08} (Figures~\ref{fig:concentration_wallA_tree}~(a) and \ref{fig:concentration_wallB_tree}~(a)) and numerical results from Merlier~\textit{et al.}~\cite{MERLIER18} (Figures~\ref{fig:concentration_wallA_tree}~(b) and \ref{fig:concentration_wallB_tree}~(b)), alongside our simulations with and without a WF.
For wall A, Figures~\ref{fig:concentration_wallA_tree}~(c) and (e) confirm that the general distribution of concentration aligns with the experimental results in (a). However, differences can be observed in the stratification, with our results showing higher concentration gradients, especially near the center of the wall. The comparison between our results with (e) and without WF (c) shows minimal variation, indicating that WF has little influence in this specific setup. 
For wall B, Figures~\ref{fig:concentration_wallB_tree}~(c) and (e) are consistent with the experimental results in (a). The concentration remains lower compared to wall A, as expected, and the inclusion of WF (e) does not significantly alter the distribution. Our results also compare well with both Merlier~\textit{et al.}~\cite{MERLIER18} (b) and Gromke~\textit{et al.}~\cite{GROMKE08} (d), indicating that the general pollutant dispersion trends are well captured by our simulations.
In summary, while the concentration gradients and layering differ slightly from the experimental data, our simulations effectively capture the key pollutant distribution trends in the presence of trees. The inclusion of WF enhances accuracy but is not essential for reproducing the general distribution. Given that WF increases computational costs, further optimizations are required before integrating it into large-scale urban digital twin applications without performance trade-offs.
\subsubsection{Discussion}
The differences between our results using the HLBM approach and those from the Reynolds Stress Model (RSM) by Gromke~\textit{et al.}~\cite{GROMKE08} for the tree-free canyon with a WF are small. However, our approach shows better alignment with the experimental results. This is because RSM relies on turbulence closure equations, which approximate turbulence rather than resolving it, whereas in this work, LES is used, which partly resolves turbulence structures. Furthermore, RSM requires the use of a WF to accurately compute near-wall turbulence, whereas the HLBM approach achieves good agreement even without a WF.
This observation is also supported by the work of Merlier~\textit{et al.}~\cite{MERLIER18}, where the differences arise due to the use of the Synthetic Eddy Method (SEM), while in this work, the vortex method is applied. This distinction is also reflected in the concentration distribution: our results without a WF closely align with those of Merlier~\textit{et al.}~\cite{MERLIER18}, while our results with a WF show strong agreement with the experimental data.
For the velocity results in a canyon with a tree, our simulations do not match the experimental results as closely as in the tree-free canyon. The key difference lies in how porosity is incorporated into the governing equations. While we use \eqref{eq:lattice_porosity}, Merlier~\textit{et al.}~\cite{MERLIER18} employs a pressure loss coefficient, and Gromke~\textit{et al.}~\cite{GROMKE08} uses the porous media model in ANSYS FLUENT, which introduces a momentum sink term to account for the resistance induced by vegetation.
For concentration distribution, the results differ significantly due to differences in diffusion, which is directly influenced by the spatial resolution used in each study. Both reference cases applied grid refinement, which was not used in our work. Consequently, the turbulent Schmidt number \(S\!c_{\mathrm{t}}\) in our setup differs (see Table~\ref{tab:numerical_parameters}), since the diffusion is resolution dependent. As a result, the diffusion of concentration in our case does not match that of the reference studies.


\section{Results of the digital twin}\label{sec:res}

The area selected for the testing of the DT prototype is part of the city of Reutlingen, Germany. The area selection is described in Section~\ref{sec:osm_data} and Figure~\ref{fig:complete_geometry}.
The area was selected due to the availability of publicly accessible air pollution data from two monitoring stations, which can be used for model updates and to evaluate the simulation results.

\subsection{Setup}

\begin{figure}[h!]
    \centering
    \begin{overpic}[width=\textwidth]{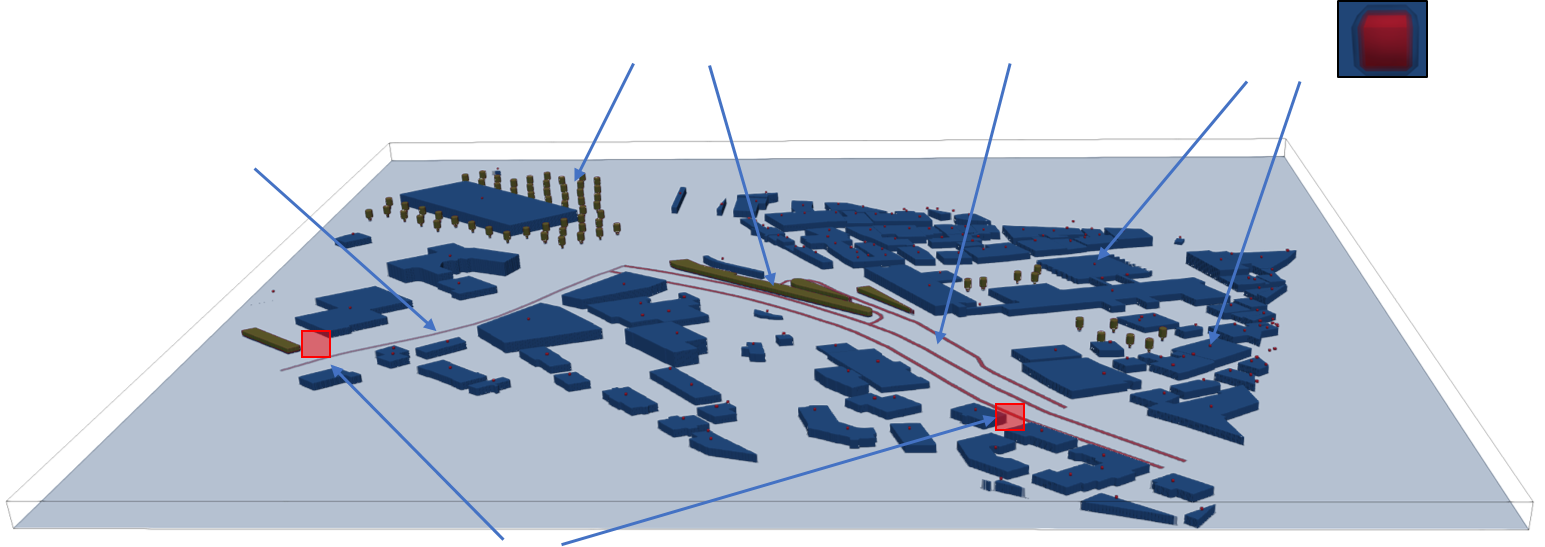}
    \put(26,-1.5){\textcolor{black}{Measuring stations}}
    \put(07,26){\textcolor{black}{Alteburgstraße}}
    \put(60,32){\textcolor{black}{Lederstraße}}
    \put(77,34){\textcolor{black}{Building}}
    \put(77,32){\textcolor{black}{exhaust}}
    \put(36,32){\textcolor{black}{Tree \& vegetation}}
    \end{overpic}
    \caption{Computational domain and measurement setup for the DT simulation. Buildings are colored blue, while trees and vegetation are colored green. Two measuring stations (red squares) are placed along Alteburgstraße (M1) and Lederstraße (M2) to monitor air pollution. The road network is colored red to indicate major traffic emission sources. Building exhaust emissions are visualized with red markers on structures, highlighting additional stationary pollution sources. The model integrates real-time data from these stations to dynamically update the simulation of pollutant dispersion and airflow behavior.}
    \label{fig:setup}
\end{figure}

To construct the DT of the urban environment, a high-resolution computational model was developed using data from OSM, meteorological stations, and emission sources as shown in Figure~\ref{fig:setup}. The selected study area includes Alteburgstraße and Lederstraße, two major roads in the region, as well as vegetation, buildings, and traffic-related emission sources. The simulation aims to model pollutant dispersion and air flow dynamics in this urban neighborhood, considering contributions from road traffic emissions, building exhaust sources, and environmental factors such as the effects of wind and vegetation.
Key sources of pollution in the study area include the following:
\begin{itemize}
    \item Road Transport Emissions: Emissions from vehicles traveling along Alteburgstraße and Lederstraße.
    \item Building Exhaust Emissions: Pollutants released from heating, ventilation, and industrial sources.
    \item Vegetation Effects: Trees and vegetation act as natural barriers that influence pollutant dispersion by affecting airflow and deposition processes.
\end{itemize}

The numerical setup is the same as in the validation from Section~\ref{sec:sim_res} with the exception being the vortex method, since it is not suitable for changing wind directions. Instead, the turbulence is induced by varying the wind direction with in a 5\% margin of the current wind direction. 
Figure~\ref{fig:conc_wind_graphs} shows the $\text{NO}_2$ and $\text{PM}_{2.5}$ concentrations of measured from the measuring station over the course of the week. It is observed that the $\text{NO}_2$ value highly correlates with the time of date, which is as expected when considering work traffic. The $\text{PM}_{2.5}$ values are more stable during the week.

\input{conc_graph}
Table~\ref{tab:numerical_parameters_dt} shows the physical and numerical parameters of the simulation. The Reynolds number was omitted because it being not practical for this simulation, because of its relation to the wind speeds. $T_{\mathrm{up}}$ represents the amount of time until the simulation is updated with new wind and concentration data.
\begin{table}[h!]
\centering
\caption{Physical and numerical parameters.}
\begin{tabular}{ccccccc}
\toprule
$S\!c_\mathrm{t}\,[-]$ & $\triangle x\,[\text{m}]$ & $\triangle t\,[\text{s}]$ & $U_\mathrm{e}\,[\text{m/s}]$ &$T_\mathrm{up}\,[\text{s}]$&$T_\mathrm{tot}\,[\text{s}]$ & \#cells$\,[-]$\\
\midrule
$0.3$  &$1$ &$1.3\times 10^{-3}$ &$0.2054$&$46000$&$0.12$&$48.25\times10^6$\\
\bottomrule
\end{tabular}
\label{tab:numerical_parameters_dt}
\end{table}

The simulation achieved similar performance results as shown in Table~\ref{tab:performance_canyon} for the configuration with trees and without WF.

\subsection{Simulation results}

\begin{figure}[h!]
\centering
    \subfloat[]{
        \begin{overpic}[width=0.48\textwidth]{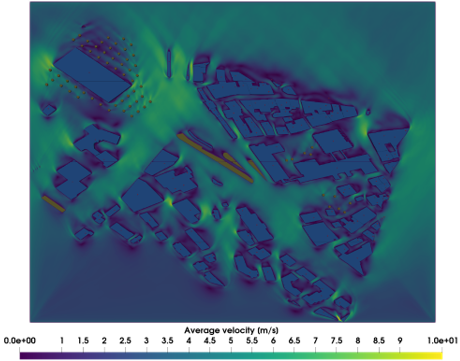}
            \put(6,9){\windrose[white]{74}{0.3}} 
        \end{overpic}
    }
    \hfill
    \subfloat[]{
        \begin{overpic}[width=0.48\textwidth]{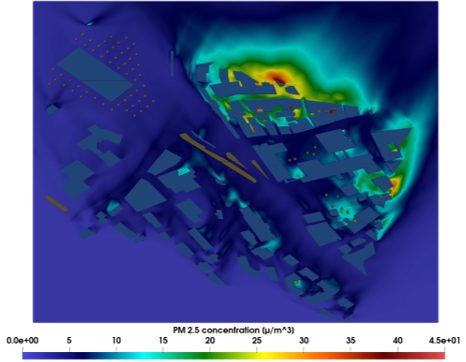}
        \end{overpic}
    }\\
    \hfill
    \subfloat[]{
        \begin{overpic}[width=0.48\textwidth]{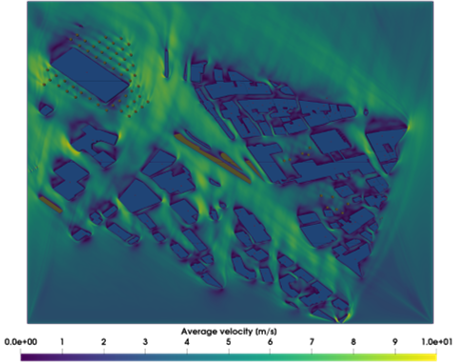}
            \put(6,9){\windrose[white]{110}{0.3}} 
        \end{overpic}
    }
    \hfill
    \subfloat[]{
        \begin{overpic}[width=0.48\textwidth]{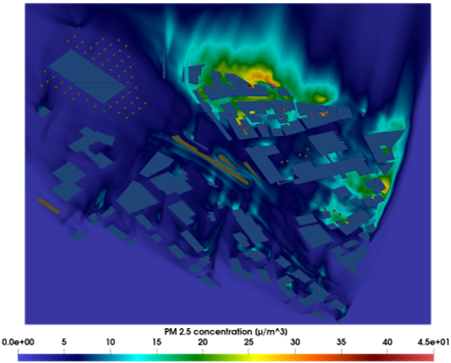}
        \end{overpic}
    }\\
    \hfill
    \subfloat[]{
        \begin{overpic}[width=0.48\textwidth]{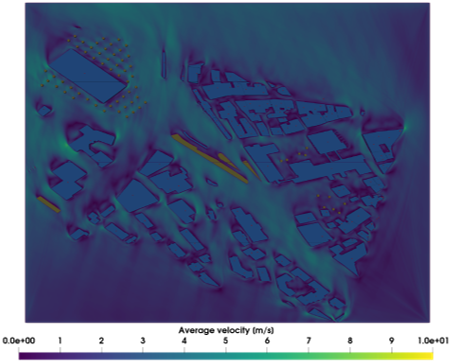}
            \put(6,9){\windrose[white]{167}{0.3}} 
        \end{overpic}
    }
    \hfill
    \subfloat[]{
        \begin{overpic}[width=0.48\textwidth]{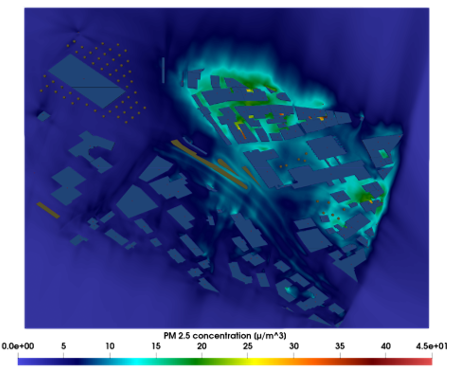}
        \end{overpic}
    }
\caption{Results of the DT from November 7 to November 13, 2024, with hourly updates. The left column (a, c and e) shows the velocity distribution, while the right column (b, d and f) presents the corresponding pollution concentration.}
\label{fig:vel_ade_dt}
\end{figure}

The DT simulation was conducted from November 7 to November 13, 2024, with hourly updates, resulting in a total of 162 updates. The results presented in Figure~\ref{fig:vel_ade_dt} illustrate the velocity distribution (left column) and the \(\text{PM}_{2.5}\) concentration distribution (right column) at different timestamps within the simulation period.

The velocity distributions shown in Figures~\ref{fig:vel_ade_dt}~(a), (c), and (e) depict the averaged wind flow at a height of 2 meters above ground level, representing pedestrian exposure. The prevailing wind direction on November 7 (Figure~\ref{fig:vel_ade_dt}~(a)) is predominantly from the northwest, leading to strong airflow through urban canyons and open spaces. By November 9 (Figure~\ref{fig:vel_ade_dt}~(c)), the wind intensity appears slightly reduced, with minor directional changes. The flow near built-up areas remains relatively stable, while turbulence effects are noticeable around obstacles. On November 13 (Figure~\ref{fig:vel_ade_dt}~(e)), the wind direction has changed to the northeast, affecting airflow patterns and creating lower-velocity zones where stagnation may occur.  

The corresponding $\text{PM}_{2.5}$ concentration distributions in Figures~\ref{fig:vel_ade_dt}~(b), (d), and (f) reveal the relationship between wind dynamics and pollutant accumulation. On November 7 (Figure~\ref{fig:vel_ade_dt}~(b)), pollution hot-spots are primarily located in the southeastern part of the domain, where lower ventilation leads to increased concentration levels. This aligns with the northwest wind direction, which carries pollutants toward these regions. By November 9 (Figure~\ref{fig:vel_ade_dt}~(d)), the pollution distribution has changed slightly due to changing meteorological conditions, yet high concentrations remain in areas with reduced airflow. On November 13 (Figure~\ref{fig:vel_ade_dt}~(f)), the northeast wind redistributes pollutants, forming new hot-spots while decreasing concentrations in previously affected areas.  

The results highlight the strong correlation between wind direction and pollutant dispersion patterns. The built-up areas and the effects of the urban canyons contribute to localized stagnation zones where pollutants remain trapped for extended periods. The formation of these hot-spots suggests the necessity for mitigation strategies such as geometrical adjustments to building layouts or the strategic placement of vegetation to enhance natural dispersion.  

The findings derived from the DT provide valuable insights for both real-time air quality assessments and long-term urban planning. By integrating simulation-based predictions with real-time data, decision-makers can develop adaptive strategies to optimize airflow and reduce pollutant exposure in critical areas, ultimately improving urban air quality and public health.  

\section{Conclusion}\label{sec:conc}
In this work, a prototype for a DT of an urban environment was developed, utilizing HLBM to model airflow and pollutant dispersion. The methodology was validated in Section~\ref{sec:sim_res}, where simulation results demonstrated strong agreement with experimental data from wind tunnel studies. The velocity and concentration distributions obtained from the DT accurately reflected expected pollutant dispersion patterns with the given geometry, confirming the feasibility of the approach.
A key advantage of this DT framework is its adaptability to different urban areas. By leveraging OSM data, the automatic geometry generation allows for rapid and efficient adaptation to new cityscapes, making the approach highly scaleable. Additionally, the DT was dynamically fed with data from meteorological measuring stations and ran over a period of one week (November 7 to November 13, 2024), enabling the observation of general pollution distribution trends in the study area.
This work serves as a proof of concept, demonstrating the potential of DTs in urban air quality monitoring. However, to further enhance the accuracy and reliability of the model, future work will involve the deployment of manually placed measuring stations throughout a city. These additional data sources will be used to fine-tune and validate the DT, improving its ability to capture real-world pollutant dispersion dynamics.
Furthermore, while the use of a WF improves near-wall flow resolution and enhances simulation accuracy, our results indicate that it is not strictly necessary to observe the general concentration distribution. Nevertheless, future efforts will focus on optimizing WF implementations to enhance computational efficiency while maintaining accuracy. The WF will be further refined and used in conjunction with the DT to achieve a more precise representation of urban flow conditions.
This study lays the foundation for a fully operational DT that can be continuously updated, validated, and refined using real-world sensor networks. In combination with high-performance numerical modeling, the DT framework will provide an essential tool for urban planners and environmental policymakers, supporting data-driven decisions aimed at improving air quality and public health in cities.

\newpage
\section*{Author contribution}\noindent
\textbf{D.\ Teutscher}:
Conceptualization,
Methodology,
Validation, 
Formal Analysis, 
Investigation,
Software,
Data Curation, 
Writing - Original Draft;
\textbf{F.\ Bukreev}: 
Writing - Review \& Editing.; 
\textbf{A.\ Kummerländer}: 
Writing - Review \& Editing.; 
\textbf{S.\ Simonis}:
Conceptualization, Methodology, Formal Analysis, Writing - Review \& Editing;
\textbf{P.\ Bächler}:
Writing - Review \& Editing;
\textbf{A.\ Rezaee}:
Writing - Review \& Editing;
\textbf{M.\ Hermansdorfer}:
Funding acquisition,
Writing - Review \& Editing;
\textbf{M.J.\ Krause}: 
Resources, 
Funding acquisition, 
Writing - Review \& Editing.
All authors have read and approved the final version of the manuscript. 

\section*{Acknowledgements}
 
The current research is part of the project "TreeCFD – Advancing Urban Sustainability through High-Fidelity Microclimate Modeling of Trees" and was partially funded by Ramboll Foundation. 
The authors gratefully acknowledge the computing time made available to them on the high-performance computer HoreKa at the NHR Center at KIT. This center is jointly supported by the Federal Ministry of Education and Research and the state governments, for HoreKa supercomputer the Ministry of Science, Research and the Arts Baden-Württemberg, participating in the National High-Performance Computing (NHR) joint funding program (\url{https://www.nhr-verein.de/en/our-partners}). 

\section*{Declaration of generative AI and AI-assisted technologies in the writing process}
During the preparation of this work, the authors used ChatGPT 4.0 in order to enhance the readability. After using this tool, the authors reviewed and edited the content as needed and take full responsibility for the content of the published article.

\section*{Data availability statement}
The numerical model has been implemented in the open-source software \textit{OpenLB} that is released under the GNU GPL v2 license. 
The latest release of \textit{OpenLB} is publicly available on Gitlab \url{https://gitlab.com/openlb/release}. 
The computationally generated data is available upon reasonable request.

\bibliographystyle{elsarticle-num} 
\bibliography{cas-refs}

\end{document}

%% file: conc_graph.tex
\begin{figure}[h!]
    \centering
    \subfloat[Concentration]{
        \begin{tikzpicture}
            \begin{axis}[
                width=7cm,
                height=6cm,
                ylabel={Concentration [$\mu\text{g}/\text{m}^3$]},
                ylabel style={yshift=-13pt},
                xticklabel style={rotate=45, anchor=east, font=\tiny},
                legend pos=north west,
                legend columns=2, 
                legend style={font=\scriptsize},
                grid=major,
                ymin=0,
                enlargelimits=true,
                xtick=data, %
                xtick={0, 16, 32, 48, 64, 80, 96, 112, 128, 144, 160},
                xticklabels={
                    07.11.2024 08:00, 
                    07.11.2024 24:00, 
                    08.11.2024 16:00, 
                    09.11.2024 08:00, 
                    09.11.2024 24:00, 
                    10.11.2024 16:00, 
                    11.11.2024 08:00, 
                    11.11.2024 24:00, 
                    12.11.2024 16:00, 
                    13.11.2024 08:00, 
                    13.11.2024 24:00
                }
            ]
                \addplot table [
                x expr=\coordindex,  
                y index=9,  
                col sep=semicolon
            ]
            {alteburgstrasse_pm25.csv};
            \addlegendentry{$\text{PM}_{2.5}$ M1}

            \addplot table [
                x expr=\coordindex,  
                y index=9,  
                col sep=semicolon
            ]
            {lederstrasse_M_pm25.csv};
            \addlegendentry{$\text{PM}_{2.5}$ M2}

            \addplot table [
                x expr=\coordindex,  
                y index=9,  
                col sep=semicolon
            ]
            {alteburgstrasse_no2.csv};
            \addlegendentry{$\text{NO}_{2}$ M1}

            \addplot table [
                x expr=\coordindex,  
                y index=9,  
                col sep=semicolon
            ]
            {lederstrasse_M_no2.csv};
            \addlegendentry{$\text{NO}_{2}$ M2}
            \end{axis}
        \end{tikzpicture}
    }
    \subfloat[Windspeed and wind direction]{
    \begin{tikzpicture}
        \begin{axis}[
            width=7cm,
            height=6cm,
            ylabel={Windspeed [$\text{m}/\text{s}$]},
            ylabel style={yshift=-13pt},
            xticklabel style={rotate=45, anchor=east, font=\tiny},
            legend pos=north west,
            legend style={font=\scriptsize},
            grid=major,
            ymin=0,
            enlargelimits=true,
            xtick=data, 
            xtick={0, 16, 32, 48, 64, 80, 96, 112, 128, 144, 160},
            xticklabels={
                07.11.2024 08:00, 
                07.11.2024 24:00, 
                08.11.2024 16:00, 
                09.11.2024 08:00, 
                09.11.2024 24:00, 
                10.11.2024 16:00, 
                11.11.2024 08:00, 
                11.11.2024 24:00, 
                12.11.2024 16:00, 
                13.11.2024 08:00, 
                13.11.2024 24:00
            }
        ]
    
            \addplot[
                color=blue,
                mark=*,
            ] table [
                x expr=\coordindex,  
                y index=7,  
                col sep=semicolon
            ] {winddatenReutlingen6_14.CSV};
            \addlegendentry{Windspeed}
    
            \addplot[
                quiver={
                    u={cos(\thisrowno{6})},  %
                    v={sin(\thisrowno{6})},  
                    scale arrows=0.5,        %
                },
                -stealth,   %
                thick,
                black,
                ] table [
                    x expr={ifthenelse(mod(\coordindex,8)==0, \coordindex, nan)}, 
                    y expr=\thisrowno{7} + 5,  %
                    col sep=semicolon
                ] {winddatenReutlingen6_14.CSV};
                \addlegendentry{Wind direction}
        \end{axis}
    \end{tikzpicture}
    }
    \caption{(a) Time series of pollutant concentrations, including $\text{PM}_{2.5}$ and $\text{NO}_2$ at two measurement stations (M1 and M2). (b) Time series of wind speed with overlaid arrows indicating wind direction. The arrows are oriented towards the cardinal directions (north, east, south, west) based on recorded wind direction data.}
    \label{fig:conc_wind_graphs}
\end{figure}

%% file: city_paper.bbl
\begin{thebibliography}{10}
\expandafter\ifx\csname url\endcsname\relax
  \def\url#1{\texttt{#1}}\fi
\expandafter\ifx\csname urlprefix\endcsname\relax\def\urlprefix{URL }\fi
\expandafter\ifx\csname href\endcsname\relax
  \def\href#1#2{#2} \def\path#1{#1}\fi

\bibitem{GRIEVES2017}
M.~Grieves, J.~Vickers, Digital Twin: Mitigating Unpredictable, Undesirable Emergent Behavior in Complex Systems, Springer International Publishing, Cham, 2017, pp. 85--113.
\newblock \href {https://doi.org/10.1007/978-3-319-38756-7\_4} {\path{doi:10.1007/978-3-319-38756-7\_4}}.

\bibitem{TAO18}
F.~Tao, J.~Cheng, Q.~Qi, M.~Zhang, H.~Zhang, F.~Sui, {Digital twin-driven product design, manufacturing and service with big data}, The International Journal of Advanced Manufacturing Technology 94 (02 2018).
\newblock \href {https://doi.org/10.1007/s00170-017-0233-1} {\path{doi:10.1007/s00170-017-0233-1}}.

\bibitem{ZHANG21}
L.~Zhang, L.~Zhou, B.~K. Horn, \href{https://www.sciencedirect.com/science/article/pii/S0278612521000455}{{Building a right digital twin with model engineering}}, Journal of Manufacturing Systems 59 (2021) 151--164.
\newblock \href {https://doi.org/https://doi.org/10.1016/j.jmsy.2021.02.009} {\path{doi:https://doi.org/10.1016/j.jmsy.2021.02.009}}.
\newline\urlprefix\url{https://www.sciencedirect.com/science/article/pii/S0278612521000455}

\bibitem{GLAESSGEN12}
E.~Glaessgen, D.~Stargel, {The digital twin paradigm for future NASA and U.S. air force vehicles}, 2012.
\newblock \href {https://doi.org/10.2514/6.2012-1818} {\path{doi:10.2514/6.2012-1818}}.

\bibitem{KLOSTERMEIER19}
R.~Klostermeier, S.~Haag, A.~Benlian, {Digitale Zwillinge – Eine explorative Fallstudie zur Untersuchung von Geschäftsmodellen}, 2019, pp. 255--269.
\newblock \href {https://doi.org/10.1007/978-3-658-26314-0\_15} {\path{doi:10.1007/978-3-658-26314-0\_15}}.

\bibitem{FACUNDO23}
F.~N. Airaudo, R.~Löhner, R.~Wüchner, H.~Antil, \href{https://www.sciencedirect.com/science/article/pii/S0045782523005959}{{Adjoint-based determination of weaknesses in structures}}, Computer Methods in Applied Mechanics and Engineering 417 (2023) 116471.
\newblock \href {https://doi.org/https://doi.org/10.1016/j.cma.2023.116471} {\path{doi:https://doi.org/10.1016/j.cma.2023.116471}}.
\newline\urlprefix\url{https://www.sciencedirect.com/science/article/pii/S0045782523005959}

\bibitem{HONGHONG23}
S.~Honghong, Y.~Gang, L.~Haijiang, Z.~Tian, J.~Annan, {Digital twin enhanced BIM to shape full life cycle digital transformation for bridge engineering}, Automation in Construction 147 (2023) 104736.

\bibitem{GAO23}
Y.~Gao, H.~Li, G.~Xiong, H.~Song, {AIoT-informed digital twin communication for bridge maintenance}, Automation in Construction 150 (2023) 104835.

\bibitem{YE19}
C.~Ye, L.~Butler, C.~Bartek, M.~Iangurazov, Q.~Lu, A.~Gregory, M.~Girolami, C.~Middleton, {A digital twin of bridges for structural health monitoring}, in: 12th International Workshop on Structural Health Monitoring 2019, Stanford University, 2019.

\bibitem{SCHROTTER20}
G.~Schrotter, C.~H{\"u}rzeler, {The digital twin of the city of Zurich for urban planning}, PFG--Journal of Photogrammetry, Remote Sensing and Geoinformation Science 88~(1) (2020) 99--112.

\bibitem{PASQUIER23}
M.~Pasquier, S.~Jay, J.~Jacob, P.~Sagaut, \href{https://www.sciencedirect.com/science/article/pii/S0360132323005899}{{A Lattice-Boltzmann-based modelling chain for traffic-related atmospheric pollutant dispersion at the local urban scale}}, Building and Environment 242 (2023) 110562.
\newblock \href {https://doi.org/https://doi.org/10.1016/j.buildenv.2023.110562} {\path{doi:https://doi.org/10.1016/j.buildenv.2023.110562}}.
\newline\urlprefix\url{https://www.sciencedirect.com/science/article/pii/S0360132323005899}

\bibitem{OpenLB}
M.~J. Krause, A.~Kummerländer, S.~J. Avis, H.~Kusumaatmaja, D.~Dapelo, F.~Klemens, M.~Gaedtke, N.~Hafen, A.~Mink, R.~Trunk, J.~E. Marquardt, M.-L. Maier, M.~Haussmann, S.~Simonis, {OpenLB—Open source lattice Boltzmann code}, Computers \& Mathematics with Applications 81 (2021) 258--288.

\bibitem{OLBRELEASE}
A.~Kummerländer, T.~Bingert, F.~Bukreev, L.~E. Czelusniak, D.~Dapelo, N.~Hafen, M.~Heinzelmann, S.~Ito, J.~Jeßberger, H.~Kusumaatmaja, J.~E. Marquardt, M.~Rennick, T.~Pertzel, F.~Prinz, M.~Sadric, M.~Schecher, S.~Simonis, P.~Sitter, D.~Teutscher, M.~Zhong, M.~J. Krause, \href{https://doi.org/10.5281/zenodo.10684609}{{OpenLB Release 1.7: Open Source Lattice Boltzmann Code}} (Feb. 2024).
\newblock \href {https://doi.org/10.5281/zenodo.10684609} {\path{doi:10.5281/zenodo.10684609}}.
\newline\urlprefix\url{https://doi.org/10.5281/zenodo.10684609}

\bibitem{USERGUIDE}
A.~Kummerländer, T.~Bingert, F.~Bukreev, L.~E. Czelusniak, D.~Dapelo, S.~Englert, N.~Hafen, M.~Heinzelmann, S.~Ito, J.~Jeßberger, F.~Kaiser, E.~Kummer, H.~Kusumaatmaja, J.~E. Marquardt, M.~Rennick, T.~Pertzel, F.~Prinz, M.~Sadric, M.~Schecher, S.~Simonis, P.~Sitter, D.~Teutscher, M.~Zhong, M.~J. Krause, \href{https://doi.org/10.5281/zenodo.13293033}{{OpenLB User Guide 1.7}} (Feb. 2024).
\newblock \href {https://doi.org/10.5281/zenodo.13293033} {\path{doi:10.5281/zenodo.13293033}}.
\newline\urlprefix\url{https://doi.org/10.5281/zenodo.13293033}

\bibitem{VANHOOFF10}
T.~{van Hooff}, B.~Blocken, \href{https://www.sciencedirect.com/science/article/pii/S1364815209001790}{{Coupled urban wind flow and indoor natural ventilation modelling on a high-resolution grid: A case study for the Amsterdam ArenA stadium}}, Environmental Modelling \& Software 25~(1) (2010) 51--65.
\newblock \href {https://doi.org/https://doi.org/10.1016/j.envsoft.2009.07.008} {\path{doi:https://doi.org/10.1016/j.envsoft.2009.07.008}}.
\newline\urlprefix\url{https://www.sciencedirect.com/science/article/pii/S1364815209001790}

\bibitem{JEANJEAN15}
A.~Jeanjean, G.~Hinchliffe, W.~McMullan, P.~Monks, R.~Leigh, \href{https://www.sciencedirect.com/science/article/pii/S135223101530248X}{{A CFD study on the effectiveness of trees to disperse road traffic emissions at a city scale}}, Atmospheric Environment 120 (2015) 1--14.
\newblock \href {https://doi.org/https://doi.org/10.1016/j.atmosenv.2015.08.003} {\path{doi:https://doi.org/10.1016/j.atmosenv.2015.08.003}}.
\newline\urlprefix\url{https://www.sciencedirect.com/science/article/pii/S135223101530248X}

\bibitem{TALEB21}
H.~M. Taleb, M.~Kayed, \href{https://www.sciencedirect.com/science/article/pii/S1618866720307329}{{Applying porous trees as a windbreak to lower desert dust concentration: Case study of an urban community in Dubai}}, Urban Forestry \& Urban Greening 57 (2021) 126915.
\newblock \href {https://doi.org/https://doi.org/10.1016/j.ufug.2020.126915} {\path{doi:https://doi.org/10.1016/j.ufug.2020.126915}}.
\newline\urlprefix\url{https://www.sciencedirect.com/science/article/pii/S1618866720307329}

\bibitem{KIM15}
K.-H. Kim, E.~Kabir, S.~Kabir, {A review on the human health impact of airborne particulate matter}, Environment international 74 (2015) 136--143.

\bibitem{BONINGARI16}
T.~Boningari, P.~G. Smirniotis, \href{https://www.sciencedirect.com/science/article/pii/S2211339816300661}{{Impact of nitrogen oxides on the environment and human health: Mn-based materials for the NOx abatement}}, Current Opinion in Chemical Engineering 13 (2016) 133--141, energy and Environmental Engineering / Reaction engineering and catalysis.
\newblock \href {https://doi.org/https://doi.org/10.1016/j.coche.2016.09.004} {\path{doi:https://doi.org/10.1016/j.coche.2016.09.004}}.
\newline\urlprefix\url{https://www.sciencedirect.com/science/article/pii/S2211339816300661}

\bibitem{DALL'OSTO13}
M.~Dall'Osto, X.~Querol, A.~Alastuey, C.~O'Dowd, R.~M. Harrison, J.~Wenger, F.~J. Gómez-Moreno, {On the spatial distribution and evolution of ultrafine particles in Barcelona}, Atmospheric Chemistry and Physics 13 (2013) 741--759.
\newblock \href {https://doi.org/10.5194/acp-13-741-2013} {\path{doi:10.5194/acp-13-741-2013}}.

\bibitem{GARCIA-MARLÈS24}
M.~Garcia-Marlès, R.~Lara, C.~Reche, N.~Pérez, A.~Tobìas, M.~Savadkoohi, D.~Beddows, I.~Samla, M.~Vörösmarty, T.~Weidinger, C.~Hueglin, N.~Mihalopolous, G.~Grivas, P.~Kalkavouras, J.~Ondracek, N.~Zikova, J.~V. Niemi, H.~E. Manninen, D.~C. Green, A.~H. Tremper, M.~Norman, S.~Vratolis, E.~Diapouli, K.~Eleftheriadis, F.~J. Gómez-Moreno, E.~Alonso-Blanco, A.~Wiedensohler, K.~Weinhold, M.~Merkel, S.~Bastian, B.~Hoffmann, H.~Altug, J.-E. Petit, P.~Acharja, O.~Favez, S.~M. Dos~Santos, J.-P. Putaud, A.~Dinoi, D.~Contini, A.~Casans, J.~A. Casquero-Vera, S.~Crumeyrolle, E.~Bourrianne, M.~Van~Poppel, F.~E. Dreesen, S.~Harni, H.~Timonen, J.~Lampilahti, T.~Petäjä, M.~Pandolfi, P.~K. Hopke, R.~M. Harrison, A.~Alastuey, X.~Querol, {Source apportionment of ultrafine particles in urban Europe}, Environment International 194 (2024) 109149.
\newblock \href {https://doi.org/10.1016/j.envint.2024.109149} {\path{doi:10.1016/j.envint.2024.109149}}.

\bibitem{DRÖGE24}
J.~Dröge, D.~Klingelhöfer, M.~Braun, D.~A. Groneberg, {Influence of a large commercial airport on the ultrafine particle number concentration in a distant residential area under different wind conditions and the impact of the COVID-19 pandemic}, Environmental Pollution 345 (2024) 123390.
\newblock \href {https://doi.org/10.1016/j.envpol.2024.123390} {\path{doi:10.1016/j.envpol.2024.123390}}.

\bibitem{MOHAN24}
V.~Mohan, V.~K. Soni, R.~K. Mishra, {Analysing the impact of day-night road traffic variation on ultrafine particle number size distribution and concentration at an urban site in the megacity Delhi}, Atmospheric Pollution Research 15 (4) (2024) 102065.
\newblock \href {https://doi.org/10.1016/j.apr.2024.102065} {\path{doi:10.1016/j.apr.2024.102065}}.

\bibitem{SAMAD22}
A.~Samad, K.~Arango, D.~A. Florez, I.~Chourdakis, U.~Vogt, {Assessment of Coarse, Fine, and Ultrafine Particles in S-Bahn Trains and Underground Stations in Stuttgart}, Atmosphere 13 (11) (2022) 1875.
\newblock \href {https://doi.org/10.3390/atmos13111875} {\path{doi:10.3390/atmos13111875}}.

\bibitem{BAECHLER24}
P.~Bächler, F.~Weis, S.~Kohler, A.~Dittler, {Exploratory measurements of ambient air quality in a residential area applying a diffusion charge based UFP monitor}, Gefahrstoffe 84 (02 2024).
\newblock \href {https://doi.org/10.37544/0949-8036-2024-01-02-17} {\path{doi:10.37544/0949-8036-2024-01-02-17}}.

\bibitem{BAECHLER21}
P.~Bächler, T.~K. Müller, T.~Warth, T.~Yildiz, A.~Dittler, {Impact of ambient air filters on PM concentration levels at an urban traffic hotspot (Stuttgart, Am Neckartor)}, Atmospheric Pollution Research 12 (6) (2021) 101059.
\newblock \href {https://doi.org/10.1016/j.apr.2021.101059} {\path{doi:10.1016/j.apr.2021.101059}}.

\bibitem{KAUR23}
K.~Kaur, K.~E. Kelly, {EPerformance evaluation of the Alphasense OPC-N3 and Plantower PMS5003 sensor in measuring dust events in the Salt Lake Valley, Utah}, Atmospheric Measurement Techniques 16 (10) (2023) 2455--2470.
\newblock \href {https://doi.org/10.5194/amt-16-2455-2023} {\path{doi:10.5194/amt-16-2455-2023}}.

\bibitem{OPENSTREETMAP}
{OpenStreetMap contributors}, \href{https://planet.openstreetmap.org/}{Planet dump}, data retrieved from OpenStreetMap and available under the Open Database License (2025).
\newline\urlprefix\url{https://planet.openstreetmap.org/}

\bibitem{TINYXML2}
L.~Thomason, {TinyXML-2}, \url{https://github.com/leethomason/tinyxml2}.

\bibitem{RIJINDERS01}
E.~Rijnders, N.~A. Janssen, P.~H. van Vliet, B.~Brunekreef, {Personal and outdoor nitrogen dioxide concentrations in relation to degree of urbanization and traffic density.}, Environmental Health Perspectives 109~(suppl 3) (2001) 411--417.
\newblock \href {https://doi.org/10.1289/ehp.01109s3411} {\path{doi:10.1289/ehp.01109s3411}}.

\bibitem{SIMONIS23}
S.~Simonis, N.~Hafen, J.~Jeßberger, D.~Dapelo, G.~Thäter, M.~J. Krause, {Homogenized lattice Boltzmann methods for fluid flow through porous media -- part I: kinetic model derivation}, ESAIM M2AN (2025).
\newblock \href {https://doi.org/10.1051/m2an/2025005} {\path{doi:10.1051/m2an/2025005}}.

\bibitem{LASAGA14}
A.~Lasaga, {Kinetic theory in the earth sciences}, Princeton University Press, 2014.

\bibitem{KUMEERLAENDER22}
A.~Kummerl{\"a}nder, F.~Bukreev, S.~F.~R. Berg, M.~Dorn, M.~J. Krause, Advances in computational process engineering using lattice boltzmann methods on high performance computers, in: W.~E. Nagel, D.~H. Kr{\"o}ner, M.~M. Resch (Eds.), High Performance Computing in Science and Engineering '22, Springer Nature Switzerland, Cham, 2024, pp. 233--247.

\bibitem{KRAUSE20171}
M.~J. Krause, F.~Klemens, T.~Henn, R.~Trunk, H.~Nirschl, {Particle flow simulations with homogenised lattice Boltzmann methods}, Particuology 34 (2017) 1--13.
\newblock \href {https://doi.org/10.1016/j.partic.2016.11.001} {\path{doi:10.1016/j.partic.2016.11.001}}.

\bibitem{JACOB18}
O.~M. Jérôme~Jacob, P.~Sagaut, {A new hybrid recursive regularised Bhatnagar–Gross–Krook collision model for Lattice Boltzmann method-based large eddy simulation}, Journal of Turbulence 19~(11-12) (2018) 1051--1076.
\newblock \href {https://doi.org/10.1080/14685248.2018.1540879} {\path{doi:10.1080/14685248.2018.1540879}}.

\bibitem{simonis2023pde}
S.~Simonis, {Lattice Boltzmann Methods for Partial Differential Equations}, Doctoral thesis, Karlsruhe Institute of Technology (KIT), {URL}: \url{https://publikationen.bibliothek.kit.edu/1000161726}. (2023).
\newblock \href {https://doi.org/10.5445/IR/1000161726} {\path{doi:10.5445/IR/1000161726}}.

\bibitem{simonis2020relaxation}
S.~Simonis, M.~Frank, M.~J. Krause, {On relaxation systems and their relation to discrete velocity Boltzmann models for scalar advection--diffusion equations}, Philosophical Transactions of the Royal Society A 378 (2020) 20190400.
\newblock \href {https://doi.org/10.1098/rsta.2019.0400} {\path{doi:10.1098/rsta.2019.0400}}.

\bibitem{simonis2022constructing}
S.~Simonis, M.~Frank, M.~J. Krause, {Constructing relaxation systems for lattice Boltzmann methods}, Applied Mathematics Letters 137 (2023) 108484.
\newblock \href {https://doi.org/10.1016/j.aml.2022.108484} {\path{doi:10.1016/j.aml.2022.108484}}.

\bibitem{GROMKE08}
C.~Gromke, R.~Buccolieri, S.~{Di Sabatino}, B.~Ruck, \href{https://www.sciencedirect.com/science/article/pii/S1352231008007553}{{Dispersion study in a street canyon with tree planting by means of wind tunnel and numerical investigations – Evaluation of CFD data with experimental data}}, Atmospheric Environment 42~(37) (2008) 8640--8650.
\newblock \href {https://doi.org/https://doi.org/10.1016/j.atmosenv.2008.08.019} {\path{doi:https://doi.org/10.1016/j.atmosenv.2008.08.019}}.
\newline\urlprefix\url{https://www.sciencedirect.com/science/article/pii/S1352231008007553}

\bibitem{MERLIER18}
L.~Merlier, J.~Jacob, P.~Sagaut, {Lattice-Boltzmann Large-Eddy Simulation of pollutant dispersion in street canyons including tree planting effects}, Atmospheric Environment 195 (2018) 89--103.
\newblock \href {https://doi.org/https://doi.org/10.1016/j.atmosenv.2018.09.040} {\path{doi:https://doi.org/10.1016/j.atmosenv.2018.09.040}}.

\bibitem{HETTEL24}
M.~Hettel, F.~Bukreev, E.~Daymo, A.~Kummerl{\"a}nder, M.~J. Krause, O.~Deutschmann, {Calculation of Single and Multiple Low Reynolds Number Free Jets with a Lattice-Boltzmann Method}, AIAA Journal (2024) 1--14.

\bibitem{GUO02}
Z.~Guo, C.~Zheng, B.~Shi, {An extrapolation method for boundary conditions in lattice Boltzmann method}, Physics of Fluids 14~(6) (2002) 2007--2010.
\newblock \href {https://doi.org/10.1063/1.1471914} {\path{doi:10.1063/1.1471914}}.

\bibitem{GUOZAHO02}
G.~Zhao-Li, Z.~Chu-Guang, S.~Bao-Chang, \href{https://dx.doi.org/10.1088/1009-1963/11/4/310}{{Non-equilibrium extrapolation method for velocity and pressure boundary conditions in the lattice Boltzmann method}}, Chinese Physics 11~(4) (2002) 366.
\newblock \href {https://doi.org/10.1088/1009-1963/11/4/310} {\path{doi:10.1088/1009-1963/11/4/310}}.
\newline\urlprefix\url{https://dx.doi.org/10.1088/1009-1963/11/4/310}

\end{thebibliography}
